\documentclass[sn-mathphys-num]{sn-jnl}
\usepackage{graphicx}%
\usepackage{multirow}%
\usepackage{amsmath,amssymb,amsfonts}%
\usepackage{amsthm}%
\usepackage{array}
\usepackage{mathrsfs}%
\usepackage[title]{appendix}%
\usepackage{xcolor}%
\usepackage{textcomp}%
\usepackage{manyfoot}%
\usepackage{booktabs}%
\usepackage{algorithm}%
\usepackage{algorithmicx}%
\usepackage{algpseudocode}%
\usepackage{listings}%
\usepackage{booktabs}

\begin{document}

\title[Article Title]{Emerging Excess Consistent with a Narrow Resonance at 152 GeV in High-Energy Proton-Proton Collisions}

\author[1,2]{\fnm{Srimoy} \sur{Bhattacharya}}\email{bhattacharyasrimoy@gmail.com}

\author[1,2]{\fnm{Benjamin} \sur{Lieberman}}\email{benjamin.lieberman@cern.ch}

\author[1]{\fnm{Mukesh} \sur{Kumar}}\email{mukesh.kumar@cern.ch}

\author[3]{\fnm{Andreas} \sur{Crivellin}}\email{andreas.crivellin@psi.ch}

\author[4]{\fnm{Yaquan} \sur{Fang}}\email{fang@mail.cern.ch}

\author[1]{\fnm{Rachid} \sur{Mazini}}\email{rachid.mazini@cern.ch}

\author[1,2]{\fnm{Bruce} \sur{Mellado}}\email{bmellado@mail.cern.ch}

\affil[1]{\orgdiv{School of Physics and Institute for Collider Particle Physics}, \orgname{University of the Witwatersrand}, \orgaddress{\street{Wits}, \postcode{2050}, \city{Johannesburg},  \country{South Africa}}}

\affil[2]{\orgdiv{iThemba LABS}, \orgname{National Research Foundation}, \orgaddress{\street{PO Box 722}, \city{Somerset West}, \postcode{7129}, \country{South Africa}}}

\affil[3]{\orgdiv{Physik-Institut}, \orgname{Universität Zürich}, \orgaddress{\street{Winterthurerstrasse 190}, \city{CH–8057 Zürich}, \country{Switzerland}}}

\affil[4]{\orgdiv{Institute of High Energy Physics}, \orgaddress{\street{19B, Yuquan Road, Shijing District}, \postcode{100049}, \city{Beijing},  \country{China}}}


\abstract{ 
The Higgs boson discovery at the Large Hadron Collider (LHC) at CERN confirmed the existence of the last missing particle of the Standard Model (SM).  The existence of new fundamental constituents of matter beyond the SM is of great importance for our understanding of Nature. In this context, indirect (non-resonant) indications for new scalar bosons were found in the data from the first run of the LHC, taken between 2010 and 2012 at CERN: an excess in the invariant mass of muon-electron pairs, consistent with a new Higgs boson ($S$) with a mass of $150\pm5$\,GeV.  Other processes with multiple leptons in the final state, moderate missing energy, and possibly (bottom quark) jets exhibit deviations from the SM predictions. These anomalies can be explained within a simplified model in which a new heavy Higgs boson $H$ decays into two lighter Higgses $S$. This lighter Higgs $S$ subsequently decays to $W$ bosons, bottom quarks and has also an invisible decay mode.

Here, we demonstrate that using this model we can identify narrow excesses in di-photon and $Z$-photon spectra around 152 GeV. By incorporating the latest measurements of di-photons in association with leptons, we obtain a combined global significance of $5.4\sigma$. This represents the highest significance ever reported for an excess consistent with a narrow resonance beyond the SM (BSM) in high-energy proton-proton collision data at the LHC. Such findings have the potential to usher in a new era in particle physics - the BSM epoch - offering crucial insights into unresolved puzzles of nature.}




\maketitle

\section{Introduction}\label{sec1}
% Intro and Higgs
The Standard Model (SM) of particle physics provides a comprehensive mathematical framework for describing the fundamental constituents of matter and their interactions at microscopic scales. The SM has undergone extensive and successful testing across a wide range of energy scales~\cite{ParticleDataGroup:2022pth} and the discovery of the Brout-Englert-Higgs boson ($h$)\cite{Higgs:1964ia,Englert:1964et} in 2012 at the Large Hadron Collider (LHC)\cite{Aad:2012tfa,Chatrchyan:2012ufa} at CERN marked the identification of its last fundamental particle. 

Furthermore, measurements of the properties of this $125$\,GeV boson agree with SM predictions~\cite{Langford:2021osp}.

Nonetheless, it is clear that the SM cannot be the ultimate theory of Nature. For example, it cannot explain the existence of Dark Matter observed at astrophysical scales nor the non-vanishing neutrino masses required by neutrino oscillation. The SM also does not explain the dominance of matter over anti-matter and suffers from internal problems such as the flavour puzzle, the hierarchy problem and the stability of the scalar potential. Interestingly, while these internal problems are related to the scalar sector, extensions of it could solve them and might as well provide explanations for the observational evidence for DM and neutrino masses. This makes the Higgs sector a particularly promising place to search for new physics.

A plethora of searches for new particles in general, and new scalars in particular, have been performed at the LHC. While these dedicated searches
have not yielded striking evidence in the accessible energy range, the so-called ``multi-lepton anomalies" have emerged. They consist of tensions in channels with multiple leptons (electrons and/or muons) in conjunction with missing energy and with or without ($b$-)jets in the final state (see Refs.~\cite{Fischer:2021sqw,Crivellin:2023zui} for reviews and references therein). These discrepancies are statistically very significant (some individually above the $5\sigma$ level~\cite{Banik:2023vxa}) and point towards the associated production of electroweak (EW) scale new scalars~\cite{vonBuddenbrock:2016rmr,Buddenbrock:2019tua,vonBuddenbrock:2020ter}. In particular, within the framework of a simplified model,  an explanation of the multi-lepton anomalies is compatible with the direct production of a scalar $H$ via gluon fusion (with a mass of $\approx 270$\,GeV) that decays dominantly into a pair of lighter scalars $S$ (one of them being off-shell) that mostly decay to $W$ bosons. For $S$, a mass of $150\pm5$\,GeV was obtained from the invariant mass of an electron muon pair stemming from non resonant $WW$ signals~\cite{vonBuddenbrock:2017gvy}.
%and top-quark differential distributions.
%  removed unpublished~\cite{Banik:2023vxa}

This suggests searching for $S$ in resonant channels in this mass region, especially in di-photon final states. In fact, in Ref.~\cite{Crivellin:2021ubm} it was shown that the side-bands of the SM Higgs boson analyses by ATLAS~\cite{ATLAS:2022tnm,Aad:2020ivc,Aad:2021qks} and CMS~\cite{Sirunyan:2021ybb,Sirunyan:2020sum,CMS:2018nlv,Sirunyan:2018tbk} suggest the presence of a narrow scalar resonance with a mass around $150$\,GeV, which is produced in association with leptons, missing energy, ($b$-)jets, $W$ and $Z$ bosons\footnote{The 152\,GeV could in particular be the neutral component of an $SU(2)_L$ triplet~\cite{Ashanujjaman:2024pky,Crivellin:2024uhc,Ashanujjaman:2024lnr} or an $SU(2)_L$ doublet~\cite{Banik:2024ugs,
Banik:2024ftv}.}, as implied by the explanation of the multi-lepton anomalies. In this study, we update the previous results regarding the narrow resonance with the mass around 150 GeV by considering the recent results reported by the LHC experiments with di-photon final states in association with leptons and taus.

% Aticle breakdown
The article is organized as follows: in \autoref{history}, we discuss the timeline of the multi-lepton anomalies and narrow excesses as the motivation for this work. To further motivate the robustness of the analysis, in \autoref{inputs} we have accumulated all the relevant channels driven by the multi-lepton anomalies and present the detailed methodology of the combination procedure that we used. Furthermore, we discuss about the important results showing a growing excesses consistent with a narrow resonance with a mass around 152\,GeV. Finally, in \autoref{conclusion}, we conclude by summarising the main results and implications of this study.

\section{Multi-lepton anomalies as indirect signs of new bosons}\label{history}

\begin{figure*}[t]
    \centering
    \includegraphics[width=0.85\linewidth]{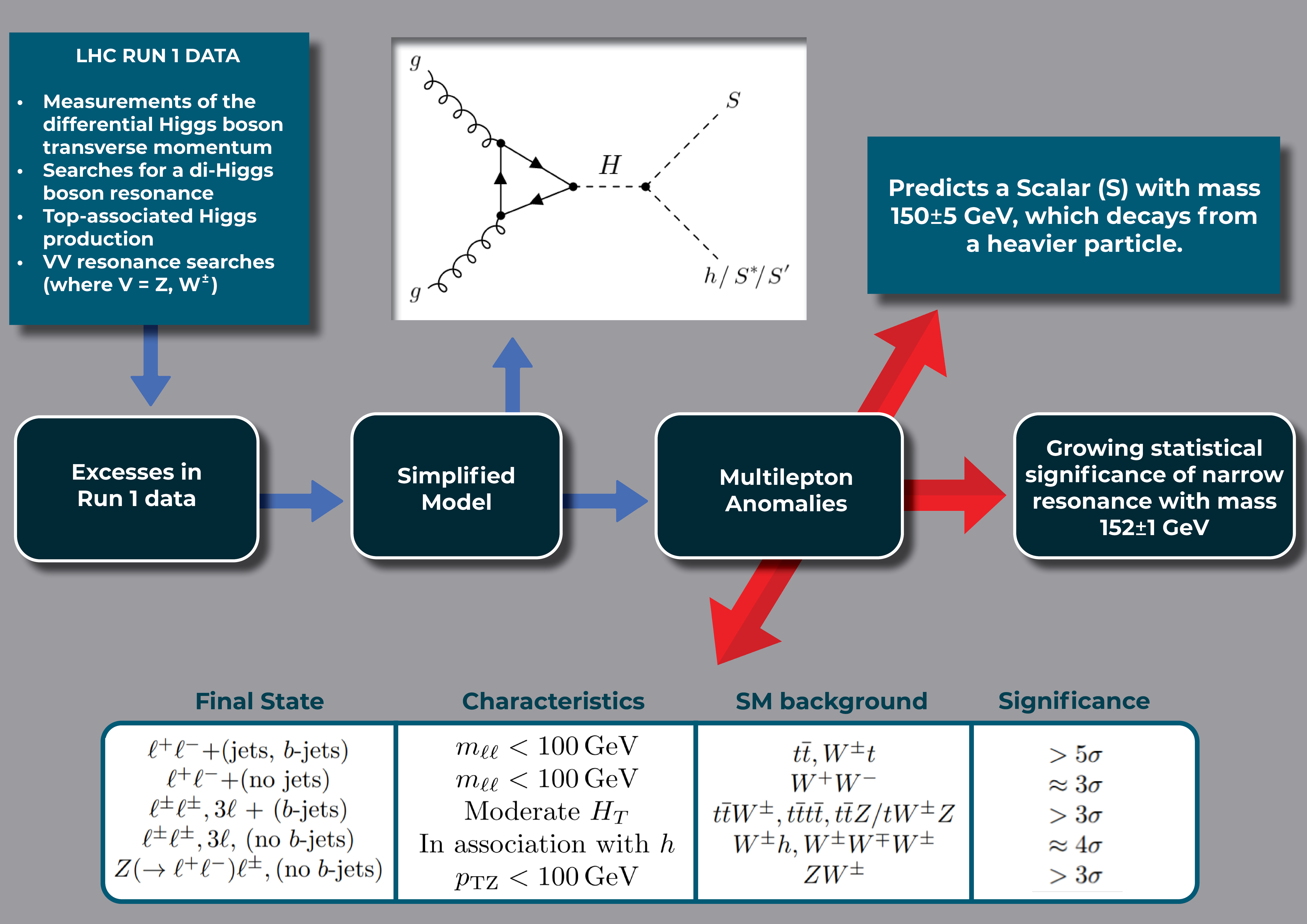}
    \caption{An overview of the timeline of the multi-lepton anomalies leading to the prediction of a narrow resonance with a mass of $150\pm5\,$GeV. The figure includes a table summarising the final states, basic characteristics, relevant SM background, and the respective significances of the multi-lepton anomalies~\cite{Crivellin:2023zui}.}
    \label{fig:modelhistory}

\end{figure*}

The discovery of the Brout-Englert-Higgs boson at the LHC completed the SM of particle physics, which has been rigorously tested and validated at both the precision and high-energy frontiers. Nevertheless, certain features and excesses observed in the LHC Run-1 data, which could indicate the presence of new scalar bosons, warrant further investigation. These findings include measurements of the differential transverse momentum of the Higgs boson ($p_{T}^h$)~\cite{ATLAS:2014yga,ATLAS:2014xzb,CMS:2015qgt,CMS:2015hja}, where $h$ denotes the SM-like Higgs boson; searches for a di-Higgs boson resonance~\cite{ATLAS:2015sxd,CMS:2014ipa,CMS:2015uzk,CMS:2014jkv}, studies of the Higgs boson in association with top quarks~\cite{ATLAS:2014ayi,ATLAS:2015xdt,ATLAS:2015utn,CMS:2014tll}, and searches for resonances decaying into two electroweak gauge boson $V V$  (where $V = Z, W^\pm$)~\cite{ATLAS:2015iie,ATLAS:2015pre,CMS:2015hra}. In 2015, a study~\cite{vonBuddenbrock:2016rmr} showed that a simplified model can be constructed to accommodate these excesses. This entailed the production of a heavy scalar, $H$, with a mass around 270\,GeV decaying into $SS,Sh,hh$, where $h$ is the Higgs boson in the SM and $S$ is a SM Higgs-like scalar as shown in ~\autoref{fig:modelhistory}.

One of the most intriguing phenomenological consequences of this model is the production of multiple leptons in association with hadronic jets. Some multi-lepton excesses, which were not initially included in the model-building procedure, were observed in the LHC Run 1 dataset. Motivated by these findings, the earlier phenomenological study was further elaborated in a subsequent analysis~\cite{vonBuddenbrock:2017gvy}, by focusing on the various multi-lepton final states. In particular, from the di-lepton component of the early multi-lepton excesses, and assuming that $S \rightarrow W^+ W^- \rightarrow \ell^+\ell^- \nu\nu$, where $\ell = e, \mu$, the mass of $S$ was determined to be $150\pm5$\,GeV. In 2017, it was decided to fix the parameters of the simplified model to the above-mentioned masses of $H$ and $S$, assuming that the heavy scalar predominantly decays into di-boson states ($Sh$). This simplified model was established prior to the release of the results from the LHC Run 2 dataset. This stringent approach was adopted to prevent fine-tuning of parameters, which could introduce biases or look-elsewhere effects, thereby compromising the calculation of the global significance of the potential excesses.

In this context, the excesses reported in Ref.~\cite{Buddenbrock:2019tua} led to the introduction of the ``multi-lepton anomalies" at the LHC. This refers to final states exhibiting statistically significant deviations from SM predictions when there are two or more leptons in the final state, both with and without $b$-jets. These discrepancies with the SM were validated by larger, independent datasets collected during LHC Run 2. This validation also included identifying excesses in multi-lepton final states that were not used for the initial model building but are predicted by the simplified model (e.g. see Ref.~\cite{Hernandez:2019geu}). Following the strict procedure outlined above and utilizing the simplified model, a combined global significance of over $8\sigma$ was reported in Ref.~\cite{Buddenbrock:2019tua}. This established that the excesses in multi-lepton final states at the LHC are not driven by statistical fluctuations. The excesses reported in Ref.~\cite{Buddenbrock:2019tua} continue to be corroborated by new datasets. For further reviews on the status of the multi-lepton anomalies at the LHC, see Refs.~\cite{Fischer:2021sqw,Crivellin:2023zui}.

~\autoref{fig:modelhistory} summarises the timeline leading to the prediction of a scalar with a mass around 150\,GeV, produced from the decay of a heavier boson. The multi-lepton anomalies observed at the LHC originated from a simplified model formulated based on early features of the LHC Run 1 data. These anomalies were further corroborated with data from LHC Run 2. Consequently, the mass of $S$ was predicted based on the multi-lepton anomalies.

In retrospect, it is noteworthy that the early features of the LHC Run 1 data, which were used to define the simplified model, emerged initially as upward statistical fluctuations. As a result, the statistical significances of these excesses have not increased proportionally to the square root of the integrated luminosity, a behavior characteristic of the initial stages of data exploration.

\section{A narrow resonant structure at 152\,GeV} \label{inputs}

As discussed in the last section, the multi-lepton anomalies predict the existence of a scalar boson with a mass of $m_S=150\pm5\,$GeV, predominantly produced in association with leptons, hadronic jets (both originating from $b$-quarks and light quarks), and missing energy (e.g., $H\rightarrow SS^{(*)}$, where the associated objects are produced via $S^{(*)}$).

Due to the potentially large number of final states, it is crucial to establish a procedure that avoids the so-called `cherry-picking,' which can introduce biases and overestimate the statistical significance. Unlike an extensive scan across the vast dataset provided by LHC experiments, the approach followed here focuses in very specific corners of the phase space, as predicted by the simplified model. We use the simplified model, whose parameters were fixed in 2017 (see above), to determine the final states that explore a narrow structure with a mass around 150\,GeV. As such, we assume that a scalar $H$ with a mass of 270\,GeV is directly produced via gluon fusion and decays into two lighter scalars $S$ ( one of them is off-shell ) that dominantly decay to $W$ bosons and $b$-jets.\footnote{References~\cite{vonBuddenbrock:2016rmr,Buddenbrock:2019tua} assumed the dominance of the $H\rightarrow Sh$ decay. While this has the same signatures as $H\rightarrow SS^*$, it leads to issues with the SM Higgs boson signal strength measurements at the LHC.}  We assume $S$ to be SM-like, meaning that it has the branching ratios of a SM-like Higgs boson with the same mass. Additionally, in Ref.~\cite{vonBuddenbrock:2016rmr}, $S$ was allowed to decay into invisible final states, leading to signatures with missing energy. This targeted strategy significantly reduces look-elsewhere effects and the associated trial factors that would stem from a general search for a resonance within the large phase space of proton-proton collisions at the LHC.

As a first step, a search for the decays $S\rightarrow\gamma\gamma, Z\gamma$ within a mass range of 130~GeV - 160\,~GeV, in association with leptons, jets, and missing energy (using the side-bands of SM Higgs analyses) was conducted in 2021~\cite{Crivellin:2021ubm}. The study yielded a global statistical significance of 3.9$\sigma$ for $m_S \sim 151.5$\,GeV. This result represents the first step towards the establishment of a narrow resonance signal using the final states predicted by the simplified model.

\begin{table*}[t]
% \begin{center}
    \centering
    \begin{tabular}{|c|c|c|}
    \hline
      2021 (Group 1)  & 2023 (Group 2)  & 2025 (Group 3)\\
    \hline         
             $\gamma \gamma + E^T_{\text{miss}}$~\cite{ATLAS:2021jbf, CMS:2018nlv}  & $\gamma \gamma + \geq (1\ell + 1 b$-jet)~\cite{ATLAS:2023omk} & $\gamma \gamma + \tau$~\cite{ATLAS:2024lhu}\\
             $\gamma \gamma + W,Z $ ~\cite{ATLAS:2020pvn, CMS:2021kom} & $\gamma \gamma + \geq $ 4 jets~\cite{ATLAS:2023omk} & $\gamma \gamma + \ell$~\cite{ATLAS:2024lhu} \\
             ${ \gamma \gamma + b\text{-jet}}$ ~\cite{ATLAS:2020ior, CMS:2020cga}  & ${ WW^{*}} + E^T_{\rm miss}$~\cite{CMS:2022uhn, ATLAS:2022ooq} & $\gamma \gamma + 2(\ell,\tau)$~\cite{ATLAS:2024lhu}\\
             $\gamma \gamma$ + $2b$-jets  ~\cite{ATLAS:2021ifb} & & \\
             $ Z(\to \ell^+ \ell^-)\gamma + 1\ell, jj$ ~\cite{CMS:2022ahq}  & & \\

    \hline

    \end{tabular}
    \vspace{0.3 cm}
    \caption{List of final states predicted by the simplified model used to interpret the corresponding experimental results. The initial combination in 2021, was done by combining all the channels mentioned in Group 1. The subsequent combination in 2023 included the channels in both Group 1 and 2. The final combination in 2025 comprehensively includes all the channels (Group 1, 2 and 3).}
    \label{tab:channels}
% \end{center}
\end{table*}

Here we present the procedure used to further search for a narrow resonance $S$ with a mass around 150\,GeV using additional data sets provided by the LHC experiments. This involves combining the experimental results  from relevant final states as input. A summary of the final states selected by the aforementioned procedure from publicly available results is provided in \autoref{tab:channels} (detailed discussion on the input channels and event selections are given in \autoref{secA1} and \autoref{secA2} respectively). 
The combination is performed in three steps:
\begin{itemize}
\item Group 1:  Here, we combine the results from LHC experiments up to 2021, where the on-shell $S$ decays into the $\gamma\gamma$ or $Z\gamma$ channels. The first combination in this mass range, presented in Ref.~\cite{Crivellin:2021ubm}, analyzed the associated production of SM gauge bosons ($\gamma\gamma$ and $Z\gamma$) with 2 DOF, reporting indications of a new scalar $S$ with a global (local) significance of $3.9\sigma (4.3\sigma)$ at 151.5 GeV. The observed yields in these channels are consistent with the predictions of the simplified model.

\item Group 2: We further consider  $\gamma \gamma + \geq(1\ell + 1 b$-jet)~\cite{ATLAS:2023omk}, $\gamma \gamma+\geq4$jets~\cite{ATLAS:2023omk},  and ${S(\to WW^{*}}) + E^T_{\rm miss}$~\cite{CMS:2022uhn, ATLAS:2022ooq} channels.

\item Group 3: Finally, we include the results reported in 2024 using the final states $\gamma \gamma + \tau$, $\gamma \gamma + \ell$ and $\gamma \gamma + 2(\ell,\tau)$~\cite{ATLAS:2024lhu}.
\end{itemize}

For each final state, we independently fit the $\gamma\gamma$ and $Z\gamma$ mass spectra and extract the p-values for masses in the range of 140~GeV - 157~GeV (see ~\autoref{secA3} for individual fits). The results for 
the ${S(\to WW^{*}}) + E^T_{\rm miss}$ final state are obtained by recasting the results from Ref.~\cite{Coloretti:2023wng}. The fiducial cross sections of individual final states are given for 152 GeV in \autoref{secA4}.
 
\begin{figure*}[t]
    \centering
    \includegraphics[width=0.9\linewidth]{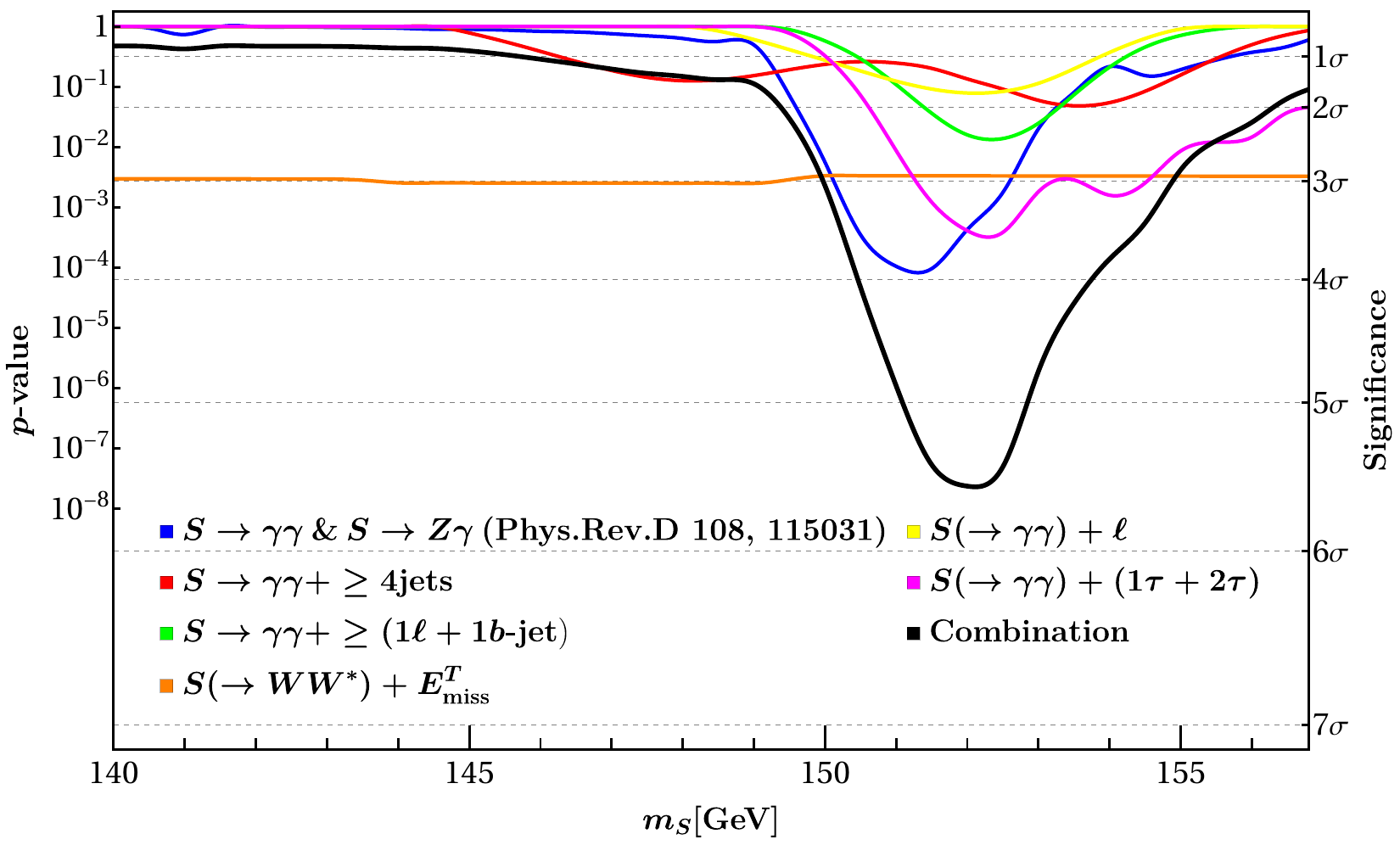}
    \caption{The $p$-values of the individual channels as well as their combination and the corresponding local significance (see text for the details). The best fit is obtained at $152$\,GeV with a global (local)  significance of 5.4$\sigma$ ($5.6\sigma$). \label{fig:Pvalue152All}}
\end{figure*}

In the final combination process, we account for the appropriate degrees of freedom (DOF) as follows. We perform a correlated combined fit of the $S(\to \gamma\gamma) + 1\tau$ and $S(\to \gamma\gamma) + 2\tau$ channels. Additionally, we include the $\gamma\gamma + 1\ell$ channel as an independent DOF to remain conservative. Thus, the final combination includes the following six DOFs:

\begin{enumerate}
    \item $S \to \gamma\gamma + Z\gamma$
    \item $S \to \gamma\gamma + \geq4$ jets
    \item $S \to \gamma\gamma + \geq(1\ell + 1b$-jet)
    \item $S \to W W^* + E^T_{\text{miss}}$
    \item $S \to \gamma\gamma + 1\ell$
    \item $S \to \gamma\gamma + (1\tau,2\tau)$ from the correlated fit.
\end{enumerate}

We combine these channels using Fisher’s method~\cite{fisher1925}, where the test statistic is given by:

\begin{equation}
    \chi^2_{2n} = -2 \sum_{i=1}^{n = 6} \log (p_i),
\end{equation}
where $p_i$ represents the $p$-value of each channel, and $\chi^2_{2n}$ follows a $\chi^2$distribution with $2n$ degrees of freedom.

Following the above-mentioned methodology, we first update the significance reported in Ref.~\cite{Crivellin:2021ubm} by incorporating results published after 2021 in the $\gamma\gamma + b$-jet~\cite{ATLAS:2021ifb} and $Z(\to \ell^+ \ell^-)\gamma + 1\ell, jj$~\cite{CMS:2022ahq} final states. The inclusion of these channels reduces the global significance from $3.9\sigma$ to $3.5\sigma$ at a mass of 152\,GeV. Next, this result is combined with the new $Z\gamma$ analysis along with the $4$jet, $WW^*+E^T_{\rm miss}$ and $\geq(1\ell+1b$-jet) channels, as explained in Ref.~\cite{Bhattacharya:2023lmu}. With this update, a global (local) significance of $4.7\sigma$ (4.9$\sigma$) with 4 DOF is found for $m_S = 151.5$\,GeV. 

Furthermore, we update the previous combination by including the results from the ATLAS collaboration on non-resonant Higgs-boson pair production in final states with leptons and taus, as mentioned in Group 3. Through a correlated combined fit of one- and two-tau final states, as illustrated in~\autoref{fig:1tau2taucombined}, we obtain a maximum local significance of $3.6\sigma$ at around 152\,GeV. While performing the correlated fit, we use the ratio of signal acceptances times efficiency for one- and two-tau final states from simulations based on the simplified model ($H \rightarrow SS^*$). By adding the $S(\to\gamma\gamma) + \ell$ fit with an additional degree of freedom (i.e., a total of 6 DOF), a global (local) significance of $5.4\sigma$ ($5.6\sigma$) is obtained for $m_S = 152$\,GeV (see the black line in \autoref{fig:Pvalue152All}). To calculate the global significance, we account for the look-elsewhere effect by introducing a trial factor of approximately 3.5~\cite{Gross:2010qma}, estimated by dividing half of the mass range by the signal resolution.

Note that the impact of experimental systematic uncertainties on statistical significance is minimal. For instance, in the discovery of the SM Higgs boson~\cite{Aad:2012tfa,ATLAS-CONF-2012-093}, their inclusion reduced the significance only slightly, from $6.0\sigma$ to $5.9\sigma$. Moreover, As shown in Ref.~\cite{Crivellin:2021ubm} the dependence of experimental systematic uncertainties on the background shape, i.e.~on spurious signals, is also very small.

\section{Conclusions}\label{conclusion}

\begin{figure*}[t]
    \centering
    \includegraphics[width=0.8\linewidth]{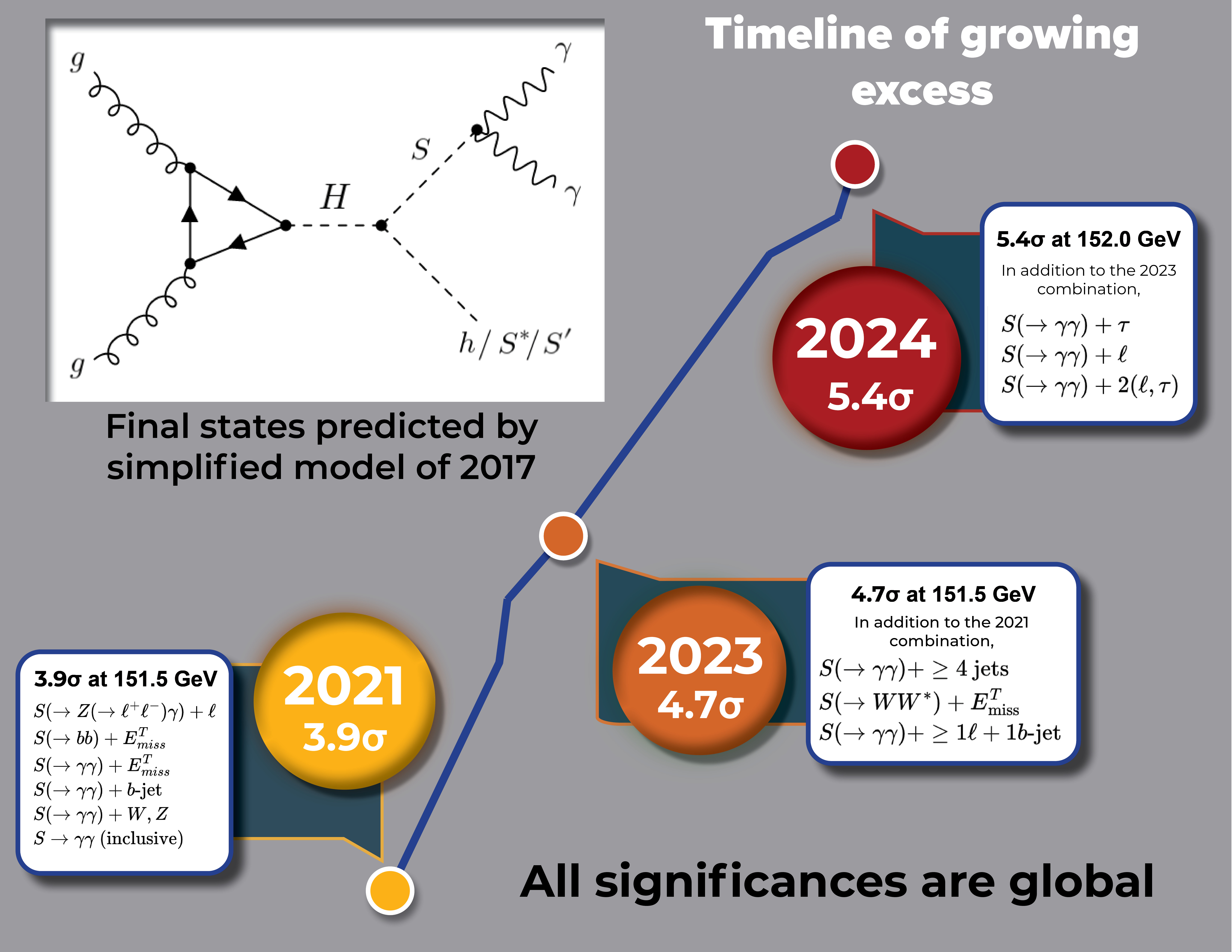}
    \caption{ Compilation of the growing excess around $152$\,GeV ordered according to the corresponding timeline. All the significances mentioned in this figure are the global significances.\label{fig:growing evidence}}
\end{figure*}

The multi-lepton excesses are very interesting in the hunt for new particles and interactions, as they are currently one of the most statistically significant deviations from SM predictions in LHC data. They can be consistently explained by assuming a simplified model in which a heavy scalar $H$ decays into two lighter scalars $(S, S^*)$ with electro-weak scale masses. With the di-lepton component of the multi-lepton anomalies particularly by the process $S\rightarrow W W^{*}\to\ell^+\ell^- + E^T_\text{miss}$, $\ell=e,\, \mu$, the simplified model predicted the mass of this scalar is to be $150\pm 5$ GeV~\cite{vonBuddenbrock:2017gvy}.

Intrigued by these excesses and their potential explanations, we searched for narrow resonances {produced in combination with leptons, ($b$-)jets and missing energy} in the side-bands of SM Higgs-boson analysis and found a possible hint for a $\approx 151$\,GeV~\cite{Crivellin:2021ubm}. Next, we added to this combination the other relevant ATLAS and CMS analyses released after 2021 and found that the significance is further strengthened: assuming a simplified model with 4 DOF ~\cite{Bhattacharya:2023lmu}. In this study, we added to this combination the latest result from ATLAS on the non-resonant Higgs-boson pair production in final states with leptons and taus, assuming the same simplified model with 6 DOF and found a global (local) significance of 5.4$\sigma (5.6 \sigma)$ for $m_S \approx 152$ GeV.

An important distinction of this study is its targeted approach to the search, which focuses on specific regions of the phase space as predicted by the simplified model, rather than scanning the entire dataset from proton-proton collisions at the LHC. This strategy effectively minimizes look-elsewhere effects and trial factors, leading to a more robust and statistically meaningful interpretation of the results.

A further confirmation of these new scalar bosons would usher in a new era in particle physics known as ``beyond-the-SM," marking one of the most substantial achievements in fundamental physics in recent decades. Additionally, a future electron-positron collider ~\cite{FCC:2018evy, CEPCStudyGroup:2018ghi} would be necessary to completely understand the characteristics and existence of these new particles, offering significant validation for their physics argument.


\begin{appendices}
%\newpage
\section{Description of input channels for the narrow resonance search}\label{secA1}

The selected input channels include all those motivated by the multi-lepton anomalies. Thus, in addition to inclusive searches, we focus on the production of $S$ in association with leptons, ($b$-)jets, and/or missing energy, while excluding other channels from the ATLAS analysis in Ref.~\cite{ATLAS:2023omk}, such as the $\gamma\gamma+\gamma$ channel, even if they show excesses. To conduct this study, we leverage CMS and ATLAS analyses of the Standard Model Higgs-boson, which inherently explore potential resonances in their side bands, covering a mass range up to 180\,GeV. However, due to some analyses ending at 160\,GeV and to avoid interference with the SM Higgs-boson resonance, we restrict our analysis to the mass range between 140\,GeV and 155-160\,GeV, as appropriate. This specific mass range is both suggested by and consistent with the observed multi-lepton anomalies. 

The description of each input channel considered in this analysis is listed below. 

$\boldsymbol{{S(\to \gamma\gamma, Z\gamma) (Phys.Rev.D\,108,\,115031)}}$: The first combination, published in Ref.~\cite{Crivellin:2021ubm}, includes data reported up until 2021, but with an update to the $Z\gamma$ channel:
\begin{itemize}
    \item ${{S(\to \gamma \gamma) + E^T_{\rm miss}}}$: In these channels, $m_S$ is reconstructed from the invariant mass of $\gamma\gamma$, and $S$ is produced in association with $E^T_\text{miss}$. The data is taken from Fig.~6 in Ref.~\cite{ATLAS:2021jbf} and Fig.~3 in Ref.~\cite{CMS:2018nlv}).

    \item ${{S(\to b \bar{b}) + E^T_{\rm miss}}}$: $m_S$ is reconstructed from the invariant mass of $b\bar{b}$, and $S$ is produced in association with $E^T_{\text{miss}}$ originating from the decay of $S$ to the invisible final states. The data is taken from Fig.~5 in Ref.~\cite{ATLAS:2020fcp}.
    
    \item ${S(\to \gamma \gamma) + b \text{-jet}}$: $m_S$ decays to two photons and is produced in association with $b$ quarks (which, in the 2HDM+S model, could originate from $S$ but also from $h$ if $H \to Sh$ is non-negligible). The data is obtained from Fig.~2 (top-right) in Ref.~\cite{ATLAS:2020ior} and Fig.~2 in Ref.~\cite{CMS:2020cga}.
    
    \item ${{S(\to \gamma \gamma) + W,Z }}$: $S$ decays to two photons and is produced in association with a $W$ or a $Z$ boson. The corresponding data is taken from Ref.~\cite{CMS:2021kom} and Fig.~9 c) and d) in Ref.~\cite{ATLAS:2020pvn}. 
    
    \item ${{S\to \gamma \gamma}}$ (inclusive): Here $m_S$ is reconstructed from the invariant mass of the photon pair while the search is quasi-inclusive. However, vector boson fusion, $W$, and $Z$ as well as top quark-associated production are excluded. Note that there is no veto on missing energy, but that this channel covers only a very tiny phase space of the quasi-inclusive final search. (see Fig.~15 (top-left) in Ref.~\cite{CMS:2021kom} and Fig.~9 a) of Ref.~\cite{ATLAS:2020pvn}).

    \item ${{S(\to Z(\to \ell^+ \ell^-)\gamma) + 1\ell, jj}}$: We use Fig.~3 top-left in Ref.~\cite{CMS:2022ahq}\footnote{Following the simplified model, we chose the two jet category that is inconsistent with the Vector Boson Fusion production mechanism.} and Fig.~3 Bottom-right in Ref.~\cite{CMS:2022ahq}. This supersedes the previously used input from Fig.~5 in Ref.~\cite{CMS:2018myz}.

    \item $S(\to\gamma \gamma)$ + 2$b$-jets: The mass of $S$ is reconstructed from the invariant mass of di-photon where $S$ is produced in association with two $b$-jets. We use Fig. 8(a) in Ref. 
    ~\cite{ATLAS:2021ifb}.
\end{itemize}

In the second combination, we have considered the following channels which published after 2021, in addition to the previously mentioned channels:

\begin{itemize}

    \item $\boldsymbol{{S(\to \gamma \gamma)}+ \geq 4j}$: Here $m_S$ corresponds to the invariant mass of the di-photon pair which is produced in association with at least 4 jets (Fig.~2 a) in Ref.~\cite{ATLAS:2023omk}). 

    \item $\boldsymbol{{S(\to WW^{*}}) + E^T_{\rm miss}}$: The CMS and ATLAS analyses of the SM Higgs-boson decaying to a pair of $W$ bosons are recast and combined. Here we use the 0-jet category for which the dominant contribution from the simplified model described above arises from $H\rightarrow S(\rightarrow WW^{*})S^{*}(\rightarrow E^T_{\rm miss})$. Other final states from associated production have very small jet veto survival probability. For ATLAS, we have used the data from Fig.~11 of Ref.~\cite{ATLAS:2022ooq} and for CMS the $m_T$ distributions ($p_{T2} < 20 \,$GeV and $p_{T2} > 20\,$GeV) of Fig.~1 of Ref.~\cite{CMS:2022uhn}. 

    \item $\boldsymbol{{S(\to \gamma \gamma)}+}\boldsymbol{ \geq (1  \ell+ 1b-}${\bf jet)}: $S$ decays to two photons and is produced in association with at least one electron or muon ($\ell$) and at least one tagged $b$-jet. The relevant experimental data are taken from Fig.~5 a) in Ref.~\cite{ATLAS:2023omk}.

\end{itemize}

Finally, in addition to these two combinations, we have considered a recent study from the ATLAS collaboration in this analysis:

\begin{itemize}
    \item $\boldsymbol{S (\to \gamma \gamma) + \ell, \tau,  2(\ell,\tau)}$: Here, the mass of $S$ is reconstructed from the invariant mass of di-photons which is produced in association with one lepton ($\mu$ or $e$) or tauon or two leptons or tauons. The relevant experimental results are taken from Fig.7 in Ref.~\cite{ATLAS:2024lhu}.

\end{itemize}

\section{Object and Event Selection}\label{secA2}

The selection criteria, applied to each channel used in this analysis, are tabulated in Table~\autoref{tab:cuts}. Here, $\gamma_1$ and $\gamma_2$ denote the leading and sub-leading photons ordered by $p_T$ respectively. The $\Delta E_T^{\text{miss}}$ is the difference between the $E_T^{\text{miss}}$ calculated from the vertex selected by the neural network and the $E_T^{\text{miss}}$ calculated from the hardest vertex. Lastly, $\Delta R \equiv \sqrt{(\Delta \eta)^2 + (\Delta \phi)^2}$ where $\phi$ is the azimuthal angle around z-axis and $\eta$ is the pseudorapidity, and $y$ is the rapidity.

\begin{center}
\begin{table}[htbp!]
   %\\centering
    \resizebox{\textwidth}{!}{%
    \begin{tabular}{c  c}
        \toprule % \hline
       Channel  & Selection Criteria \\
       \midrule % \hline
                                       & $N_\tau$ = 1, $N_\gamma$ = 2,  $N_{b-\text{jet}} = 0$, $E^{\text{miss}}_T > $35\,GeV, \\
    $S (\to \gamma \gamma) + \tau$ \cite{ATLAS:2024lhu} & 105\,GeV $>$ $m_{\gamma\gamma}$ $>$ 160\,GeV \\
     & $E_T(\gamma_1)$ $>$ 35\,GeV,  $\gamma_1 : $                                                      $p_T/m_{\gamma\gamma}$ $>$ 0.35, \\
     & $\gamma_2 : $        
                                       $p_T/m_{\gamma\gamma}$ $>$ 0.25 \\
       \midrule % \hline
       
                                       & $N_\tau$ = 0, $N_\gamma$ = 2,  $N_{b-\text{jet}} = 0$ \\
    $S (\to \gamma \gamma) + \ell$ \cite{ATLAS:2024lhu}    & $E^{\text{miss}}_T > $35 GeV, $\gamma \gamma + e 
                                       E_T(\gamma_1)$ $>$ 35\,GeV,\\
                                       & 105\,GeV $>$ $m_{\gamma\gamma}$ $>$ 160\,GeV, $\gamma_1 : $                                       $p_T/m_{\gamma\gamma}$ $>$ 0.35, $\gamma_2 : $ $p_T/m_{\gamma\gamma}$ $>$ 0.25  \\
        \midrule % \hline                         
                                       & $N_\ell + N_\tau$ = 2, $N_\gamma$ = 2,  $N_{b-\text{jet}} = 0$ \\
    $S (\to \gamma \gamma) + 2(\ell,\tau)$ \cite{ATLAS:2024lhu}  & $E^{\text{miss}}_T > $35\,GeV, 
                                       $E_T(\gamma_1)$ $>$ 35\,GeV, $m_{2(\ell,\tau)} > 12$\,GeV,\\
                                       & 105\,GeV $>$ $m_{\gamma\gamma}$ $>$ 160\,GeV, $\gamma_1 : $    $p_T/m_{\gamma\gamma}$ $>$ 0.35, $\gamma_2 : $                 $p_T/m_{\gamma\gamma}$ $>$ 0.25  \\

        \midrule % \hline                         
                                       &  $N_\gamma$ = 2,  $N_{b-\text{jet}} = 2$, $N_{\ell = e,\mu} =$ 0, \\
    $S (\to \gamma \gamma) + b \overline{b} $ ~\cite{ATLAS:2021ifb}  & $E^{\text{miss}}_T > $35\,GeV, 
                                       $E_T(\gamma_1)$ $>$ 35\,GeV, $m_{2(\ell,\tau)} > 12$\,GeV,\\
                                       & 105\,GeV $>$ $m_{\gamma\gamma}$ $>$ 160\,GeV, $\gamma_1 : $    $p_T/m_{\gamma\gamma}$ $>$ 0.35, $\gamma_2 : $                 $p_T/m_{\gamma\gamma}$ $>$ 0.25  \\
                                       &Fewer than six central ($|\eta| < 2.5$) jets are present.\\
        \midrule % \hline
                                       & $N_{\ell = e,\mu} \geq$ 1, $N_{b-\text{jet}}\geq$ 1,\\
                                       & $\gamma : p_T >$ 22\,GeV, $|\eta| <$ 2.37, $1.37 < |\eta| < 1.52$ \\
                                       & $e : p_T >$ 10\,GeV, $|\eta| <$ 2.47, $1.37 < |\eta| < 1.52$ \\
    $S (\to \gamma \gamma) + \geq 1\ell + 1 b-$jet \cite{ATLAS:2023omk} 
                                       & $\mu : p_T >$ 10\,GeV, $|\eta| <$ 2.7, 105\,GeV $>$ $m_{\gamma\gamma}$ $>$ 160\,GeV,\\
                                      & $\gamma_1 : $ $p_T >$ 35\,GeV, $p_T/m_{\gamma\gamma}$ $>$ 0.35, $\gamma_2 : $ $p_T >$ 25\,GeV, $p_T/m_{\gamma\gamma}$ $>$ 0.25 \\

        \midrule % \hline
                                       & $N_{\text{jet}} \geq$ 4, $|N_{\text{jet}}| <$ 2.5,\\
                                       & $\gamma : p_T >$ 22\,GeV, $|\eta| <$ 2.37, $1.37 < |\eta| < 1.52$ \\
    $S (\to \gamma \gamma) + \geq $ 4 jet \cite{ATLAS:2023omk} 
                                       & 105\,GeV $>$ $m_{\gamma\gamma}$ $>$ 160\,GeV, jet : $p_T >$ 25\,GeV,\\
                                       & $\gamma_1 : $ $p_T >$ 35\,GeV, $p_T/m_{\gamma\gamma}$ $>$ 0.35,\\
                                       & $\gamma_2 : $ $p_T >$ 25\,GeV, $p_T/m_{\gamma\gamma}$ $>$ 0.25\\

        \midrule % \hline
                                       & $\gamma :  |\eta| <$ 2.5 , $p_T >$ 15\,GeV, $p_T/m_{\ell^+\ell^-\gamma}$ $>$ 0.14,\\
                                       & $e :$ $p_T >$ 25 (15)\,GeV for leading (subleading),  $|\eta| <$ 2.5, \\
    $S(\to Z(\to \ell^+ \ell^-)\gamma) + 1\ell, jj$~\cite{CMS:2022ahq}
                                       & $\mu :$ $p_T >$ 20 (10)\,GeV for leading (subleading), $|\eta| <$ 2.4, \\
                                       & jet : $p_T >$ 30\,GeV, $|\eta| <$ 4.7\\
                                       & 105\,GeV $>$ $m_{\ell^+ \ell^-\gamma}$ $>$ 170\,GeV, \\
                                       &  $m_{\ell^+ \ell^-\gamma} + m_{\ell^+ \ell^-}$ $>$ 185\,GeV, \\
                                       &  $m_{\ell^+ \ell^-} >$ 50\,GeV \\

        \midrule % \hline

                                       & $\gamma :  |\eta| <$ 2.37 ,  $1.37 < |\eta| < 1.52$,\\
    $S(\to \gamma \gamma) + E^T_{\text{miss}}$ (ATLAS)~\cite{ATLAS:2021jbf}
                                       & $\gamma_1 :$ $E_T/m_{\gamma\gamma}$ $>$ 0.35,  $\gamma_2 : $ $E_T/m_{\gamma\gamma}$ $>$ 0.25,\\
                                       & 105\,GeV $>$ $m_{\gamma\gamma}$ $>$ 160\,GeV, \\
                                       & $E_T^{\text{miss}}$ $>$ 90\,GeV, $\Delta E_T^{\text{miss}} < $ 30\,GeV\\
    \cmidrule{2-2}
                                       & $\gamma :  |\eta| <$ 1.44 or $1.57 < |\eta| < 2.50$,\\
    $S(\to \gamma \gamma) + E^T_{\text{miss}}$ (CMS)~\cite{CMS:2018nlv}
                                       & $\gamma_1 :$ $p_T$ $>$ 30\,GeV, $\gamma_2 : $ $p_T$ $>$ 20\,GeV, \\
                                       & $m_{\gamma \gamma}$ $>$ 95\,GeV\\
        \midrule % \hline

                                       & Photons, Electrons, or neutral hadrons with $p_T >$ 15\,GeV,\\ 
                                       & No more than two charged particles $(p_T > 1.6$\,GeV and $|\eta| < 2.2)$,\\
                                       & $\gamma_1 :$ $p_T$ $>$ 35\,GeV, $\gamma_2 : $ $p_T$ $>$ 25\,GeV,\\
    $S(\to \gamma \gamma) + W,Z $ (CMS)~\cite{CMS:2021kom}                                    
                                       & $\gamma : $ $|\eta|$ $<$ 2.5\,GeV, except 1.44 $<|\eta|< 1.57$, \\
                                       & jets $: $ $p_T$ $>$ 25\,GeV, $|\eta|< 4.7$, \\
                                       & $\Delta$R(jet,$\gamma$) $>$ 0.4,  100\,GeV $<$ $m_{\gamma \gamma}$ $<$ 180\,GeV \\
    \cmidrule{2-2}

                                       & $\gamma : $ $|\eta|$ $<$ 2.37\,GeV, except 1.37 $<|\eta|< 1.52$,\\
                                       & $\gamma_1 :$ $p_T/m_{\gamma\gamma}$ $>$ 0.35,  $\gamma_2 : $ $p_T/m_{\gamma\gamma}$ $>$ 0.25, \\
                                       &  jets : $p_T >$ 25\,GeV, $|y| < 4.4$,  e : $p_T >$ 10\,GeV, $|\eta| < 2.47$ \\
                                       &  $\mu$ : $p_T >$ 10\,GeV, $|\eta| < 2.7$ \\
                                       & Category 1: $p_T^{W,Z} < $ 75\,GeV\\            
    $S(\to \gamma \gamma) + W,Z $ (ATLAS)~\cite{ATLAS:2020pvn}                                                              & Category 2: 75\,GeV $\leq p_T^{W,Z} < $ 150 GeV\\
                                       & Category 3: 150\,GeV $< p_T^{W,Z} < $ 250\,GeV with 0 -jet or $\geq$ 1 -jet\\
                                       & Category 4: $p_T^{W,Z} > $ 250\,GeV\\
                                       & 105\,GeV $>$ $m_{\gamma\gamma}$ $>$ 160\,GeV, \\
                                       & $\Delta R (e,\gamma \gamma) > 0.4$, $\Delta R (\text{jets},\gamma \gamma) > 0.4$, \\
                                       & $\Delta R (\text{jets},e) > 0.2$, $\Delta R (\mu,\gamma \gamma) > 0.4$, $\Delta R (\mu,\text{jets}) > 0.4$ \\
        \midrule % \hline
                                       & $\gamma_1 :$ $p_T$ $>$ 35\,GeV, $p_T/m_{\gamma\gamma}$ $>$ 0.35, \\
                                       &$\gamma_2 : $ $p_T$ $>$ 25\,GeV, $p_T/m_{\gamma\gamma}$ $>$ 0.25\\
    ${{S(\to \gamma \gamma) \geq 1t}}$ (ATLAS)~\cite{ATLAS:2020ior}
                                       & 105\,GeV $>$ $m_{\gamma\gamma}$ $>$ 160\,GeV,\\
                                       & $N_{\text{jet}}\geq 1$, $p_T^{\text{jet}} > 25$\,GeV, containing 1 $b$-jet\\
                                       & Region 1 (``Lep" region) : $N_{e,\mu} \geq$ 1, $p_T^{e,\mu} > $ 15\,GeV, \\
                                       & Region 2 (``Had" region) :  at
                                       least two additional jets \\
                                       & with $p_T > 25$\,GeV and no selected lepton\\
    \cmidrule{2-2}
                                       & $\gamma_1 :$ $p_T/m_{\gamma\gamma}$ $>$ 1/3, $\gamma_2 : $ $p_T/m_{\gamma\gamma}$ $>$ 1/4\\
    ${{S(\to \gamma \gamma) \geq 1t}}$ (CMS)~\cite{CMS:2020cga}
                                       & 100\,GeV $>$ $m_{\gamma\gamma}$ $>$ 180\,GeV,\\
                                       & Region 1 (``Lep" region) : $N_{\text{jet}}\geq 1$, $p_T^{\text{jet}} > 25$\,GeV,\\
                                       &$|\eta| < 2.4$, $N_{e,\mu} \geq$ 1, $p_T^{e} > $ 10\,GeV, $p_T^{\mu} > $ 5\,GeV,\\
                                       & Region 2 (``Had" region) :  at
                                       least 3 jets, \\
                                       & at least 1 $b$-jet and no selected lepton\\
        \bottomrule % \hline

    \end{tabular}
    }
    \caption{Selection criteria applied to each channel to form the signal pre-selection regions.  }
    \label{tab:cuts}
\end{table}
\end{center}

\section{Individual Fits for the input channels in search of the narrow resonance}\label{secA3}

In this section, we present the individual fits of the different channels considered in this analysis. We analyze the CMS and ATLAS studies to search for new scalars in the mass range between $140$\,GeV and $170$\,GeV. In each category, we model the background using the background function:
\begin{equation}
    f(m;b,\{a\}) = (1 - m)^b (m)^{a_0 + a_1 \log(m)}\,,
    \label{background}
\end{equation}
where $a_{0,1}$ and $b$ are free parameters (different for each category) and $m$ is the invariant mass of the distribution, e.g.~the di-photon mass. The choice of this particular functional form to model the background is not important for our study.~\cite{Crivellin:2021ubm} 

We incorporate a double-sided Crystal Ball function to the background, parameterized by~\autoref{background}, to model the signal contribution:
% We add to the background parameterized by Eq.~(\ref{background}) a double-sided-crystal-ball function:
\begin{equation}
   N\cdot %\left\{
   \begin{cases}
   e^{-t^{2}/2} & \mbox{if $-\alpha_{\rm Low} \leq t \leq \alpha_{\rm High}$}\\
   \frac{ e^{-0.5\alpha_{\rm Low}^{2} }}{ \left[\frac{\alpha_{\rm Low}}{n_{\rm Low}} \left(\frac{n_{\rm Low}}{\alpha_{\rm Low}} - \alpha_{\rm Low} -t \right)\right]^{n_{\rm Low}} } & \mbox {if $t < -\alpha_{\rm Low} $}\\
   \frac{ e^{-0.5\alpha_{\rm High}^{2} }}{ \left[\frac{\alpha_{\rm High}}{n_{\rm High}} \left(\frac{n_{\rm High}}{\alpha_{\rm High}} - \alpha_{\rm High}  + t \right)\right]^{n_{\rm High}} } & \mbox {if $t > \alpha_{\rm High}$}.
   \end{cases}
   \label{eq:DSCB}
\end{equation}
Here $N$ is a normalization parameter, $t = (m - m_{S})/\sigma_{CB}$ where $\sigma_{CB}$ is the width of the Gaussian part of the function, $m$ is the invariant mass of the distribution and $m_S$ the mass of the new scalar.

\begin{figure*}[ht!]
    \centering
    {\includegraphics[width=0.40\textwidth]{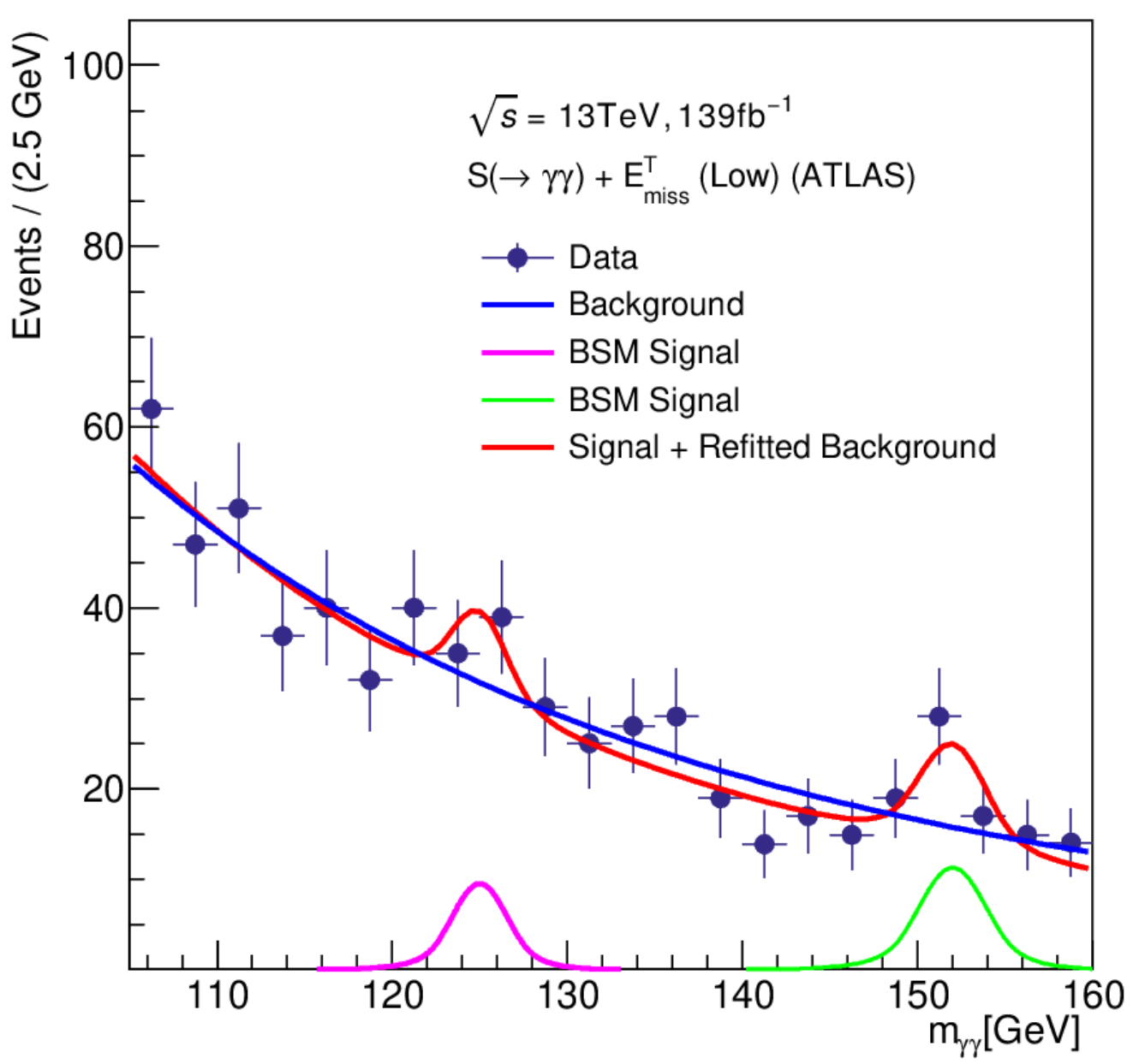}
    \label{fig:fig5b}}~~
    {\includegraphics[width=0.40\textwidth]{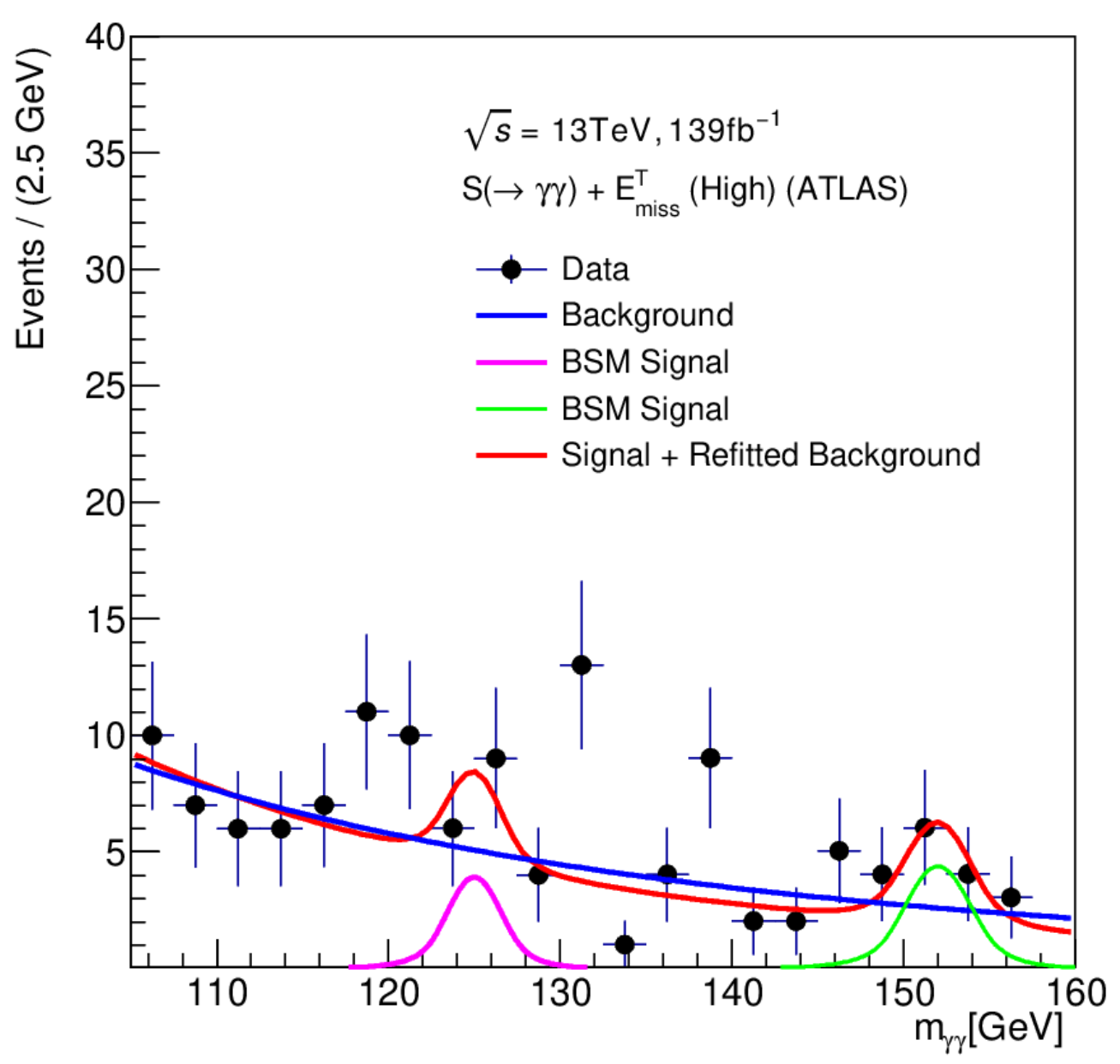}
    \label{fig:fig5b}}\\
    {\includegraphics[width=0.40\textwidth]{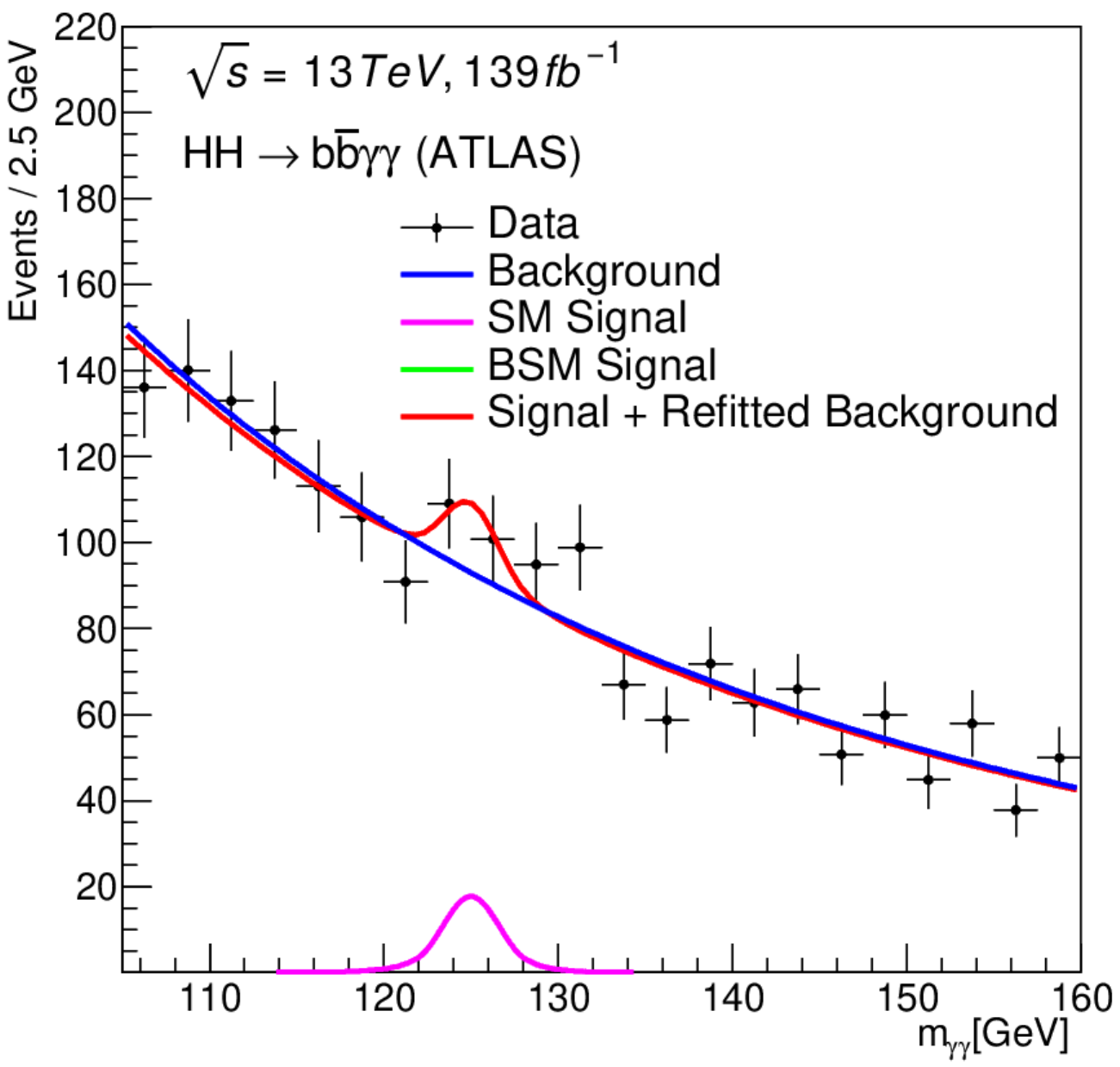}
    \label{fig:fig5c}}~~
    {\includegraphics[width=0.40\textwidth]{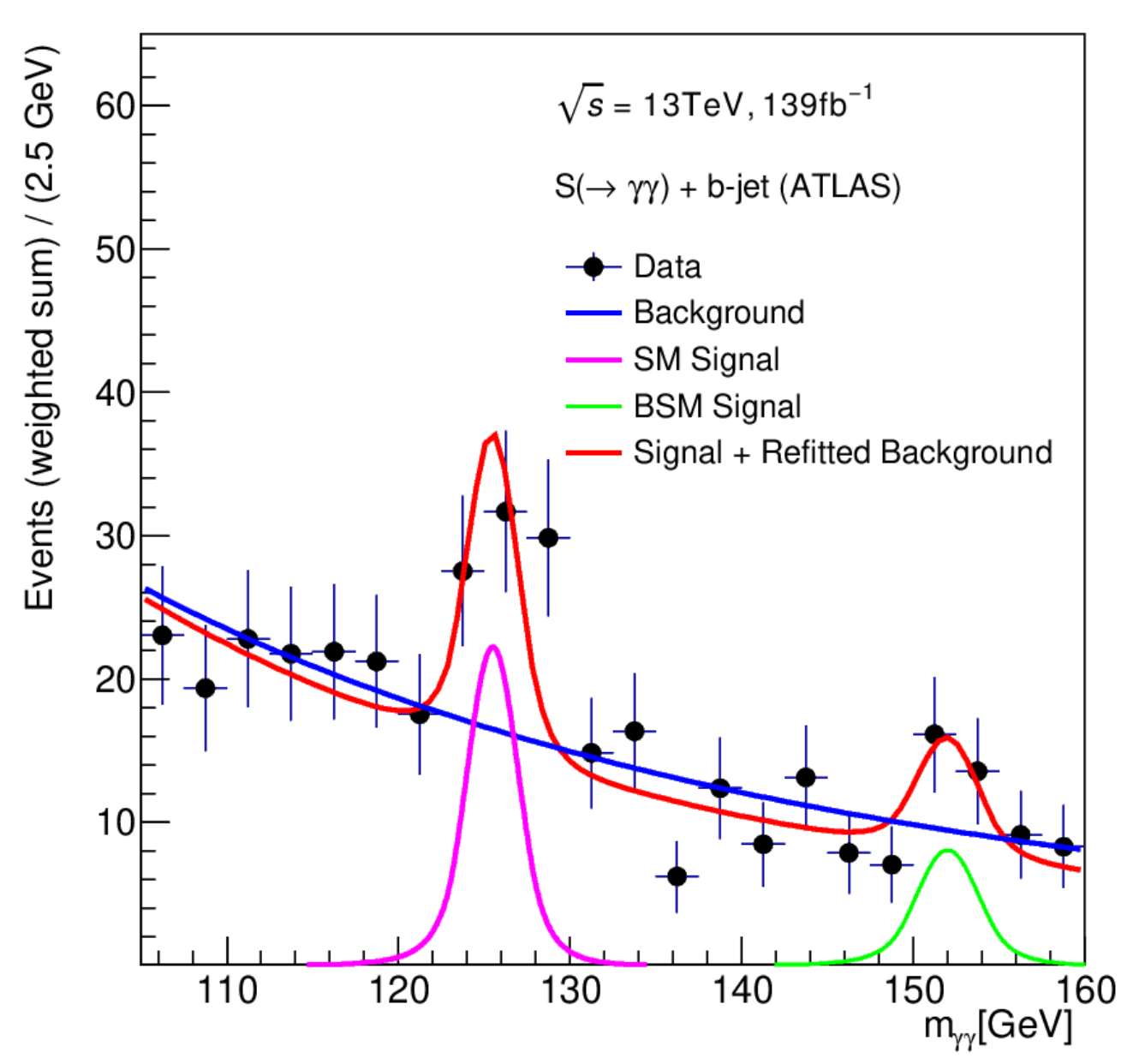}
    \label{fig:fig5d}}\\
    {\includegraphics[width=0.40\textwidth]{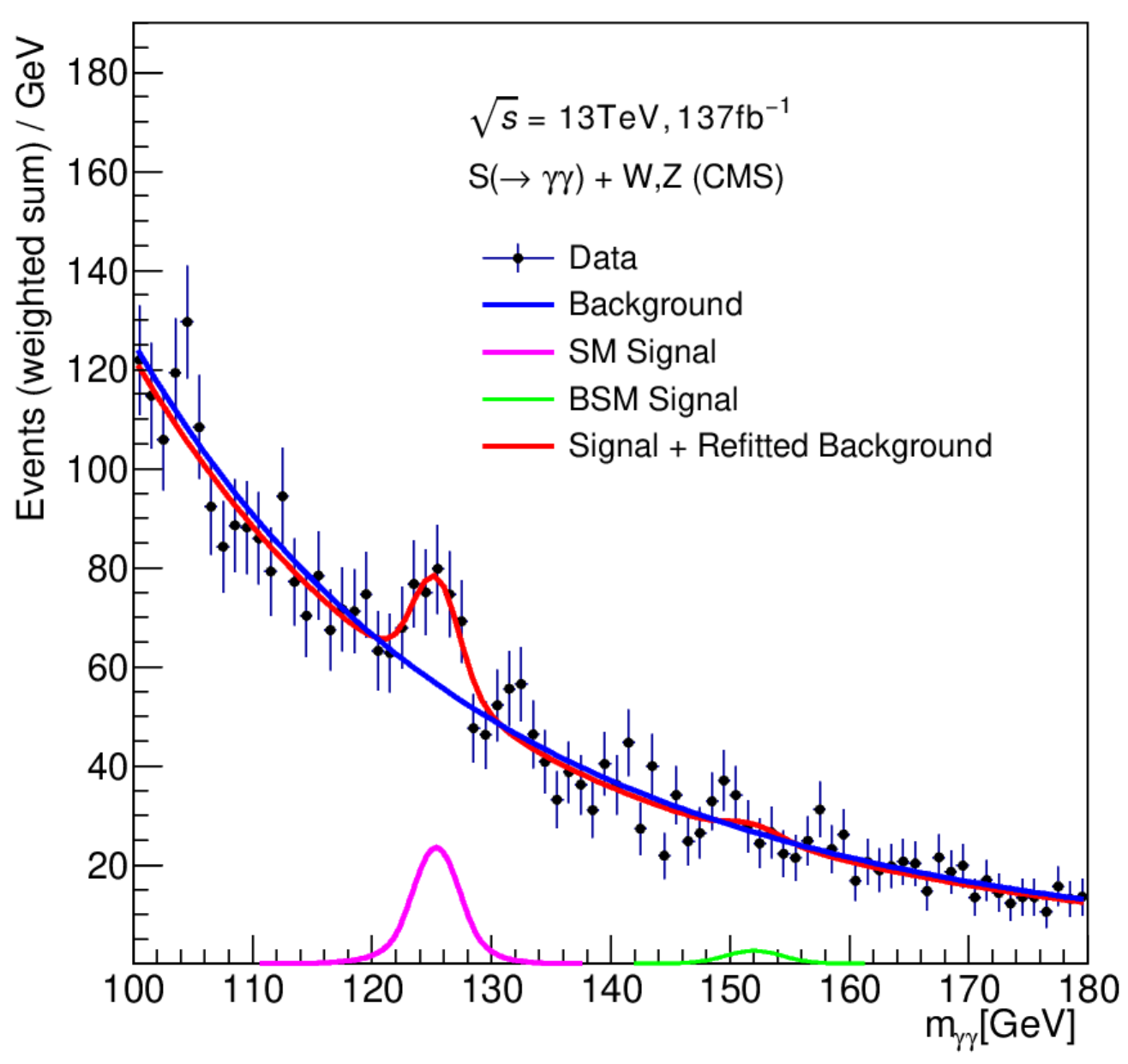}
    \label{fig:fig5e}}~~
    {\includegraphics[width=0.40\textwidth]{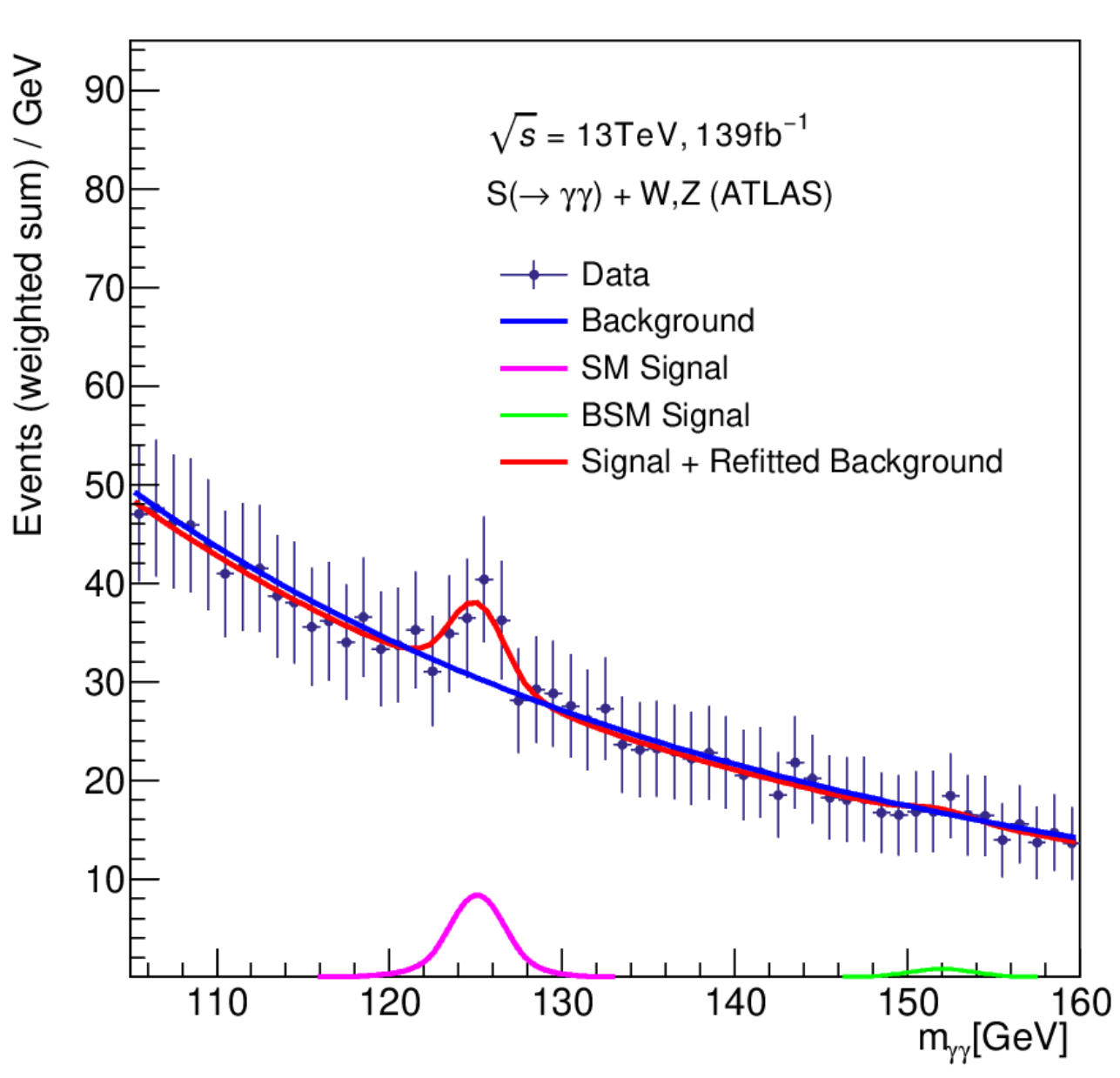}
    \label{fig:fig5f}}
    \caption{Diagrams showing the fit to background obtained within the SM, the SM Higgs signal, and the NP signal with the refitted background for five different categories. The data displayed in the last three plots corresponds to a weighted sum (see refs.~\cite{Aad:2020ivc,Sirunyan:2021ybb,ATLAS:2020pvn} for further details).}
    \label{fig:fits1}
\end{figure*}

\begin{figure*}[htbp!]
    \centering
    \includegraphics[width=0.4\textwidth]{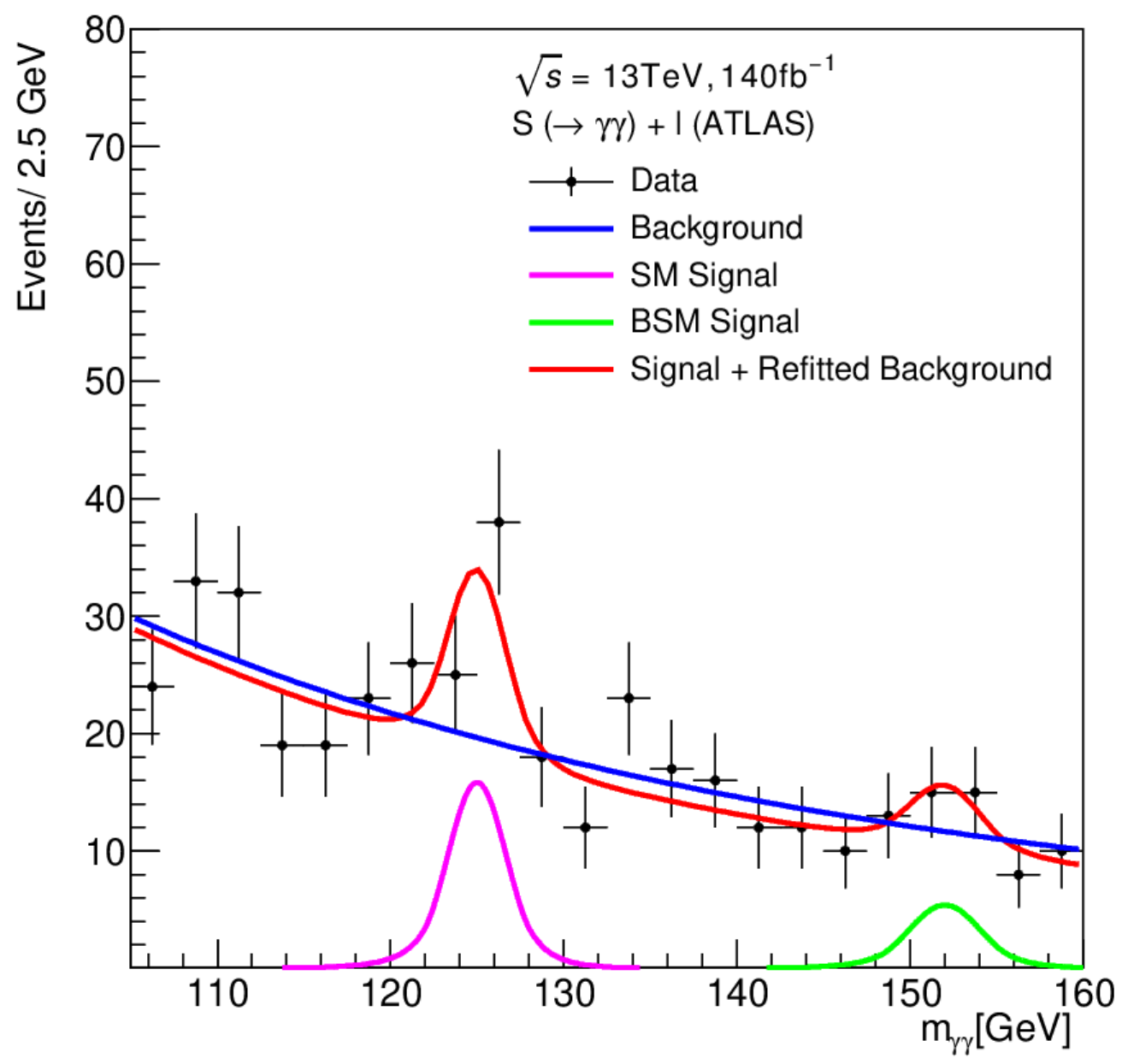}~~
    \includegraphics[width=0.4\textwidth]{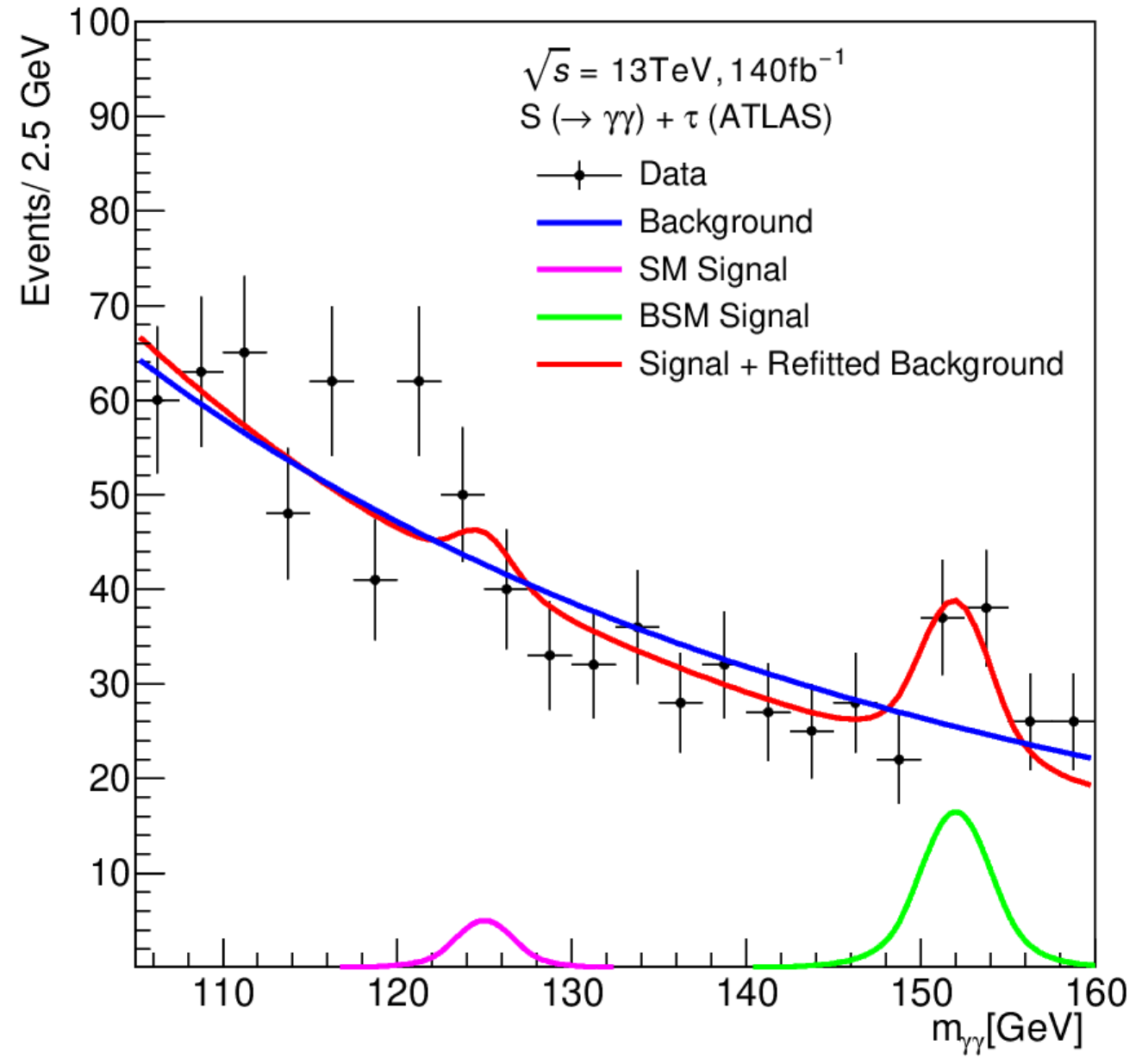}
    \includegraphics[width=0.44\textwidth]{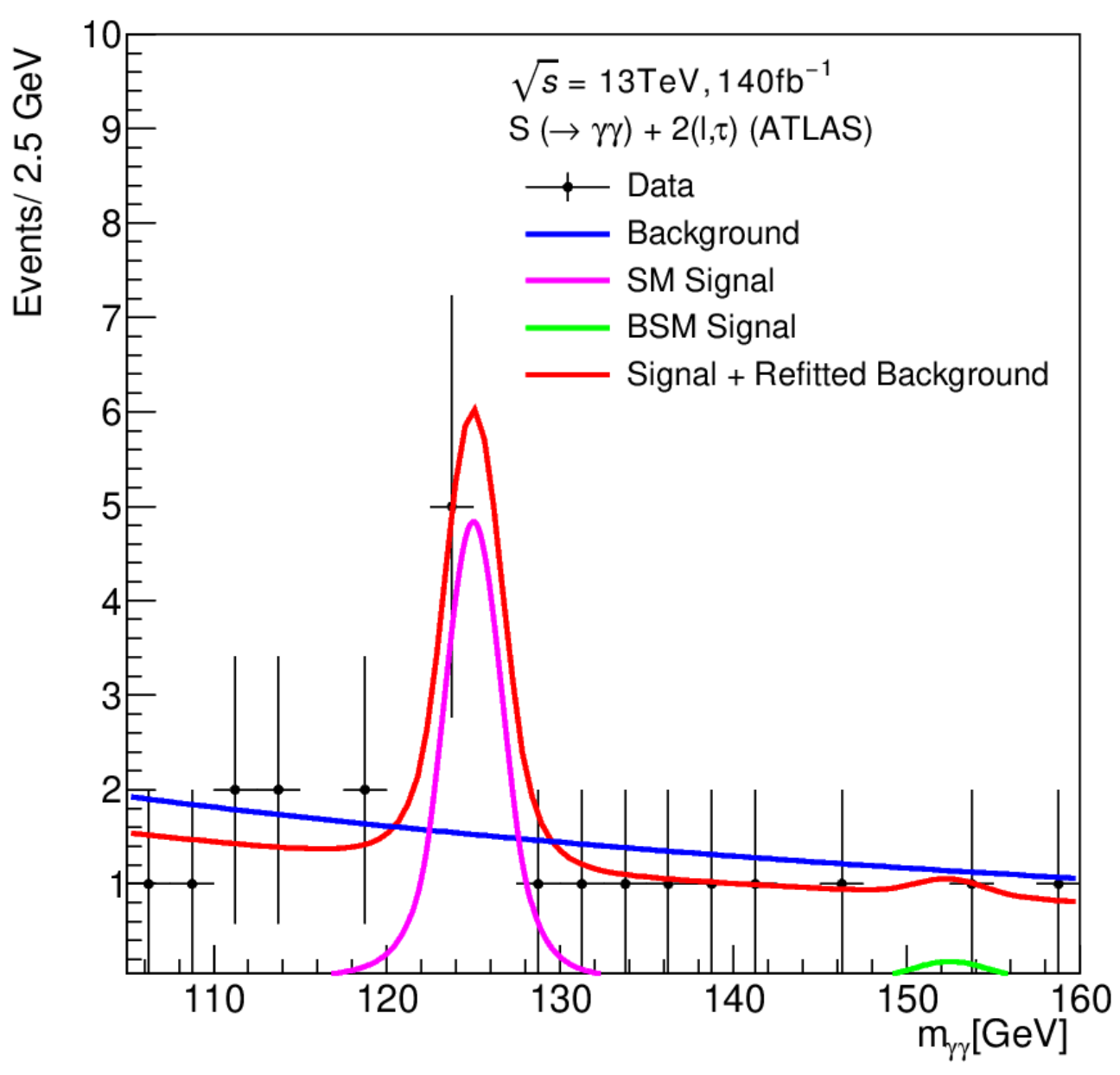}
    \caption{Diagrams showing the fit to background obtained within the SM, the SM Higgs signal, and the NP signal with the refitted background for $S(\rightarrow \gamma\gamma) + \ell$, $S(\rightarrow \gamma\gamma) + \tau$ and $S(\rightarrow \gamma\gamma) + 2(\ell,\tau)$ categories from ref.\cite{ATLAS:2024lhu}.}
    \label{fig:invmass1}
\end{figure*}

\begin{figure*}[htbp!]
    \centering
    \includegraphics[width=0.4\textwidth]{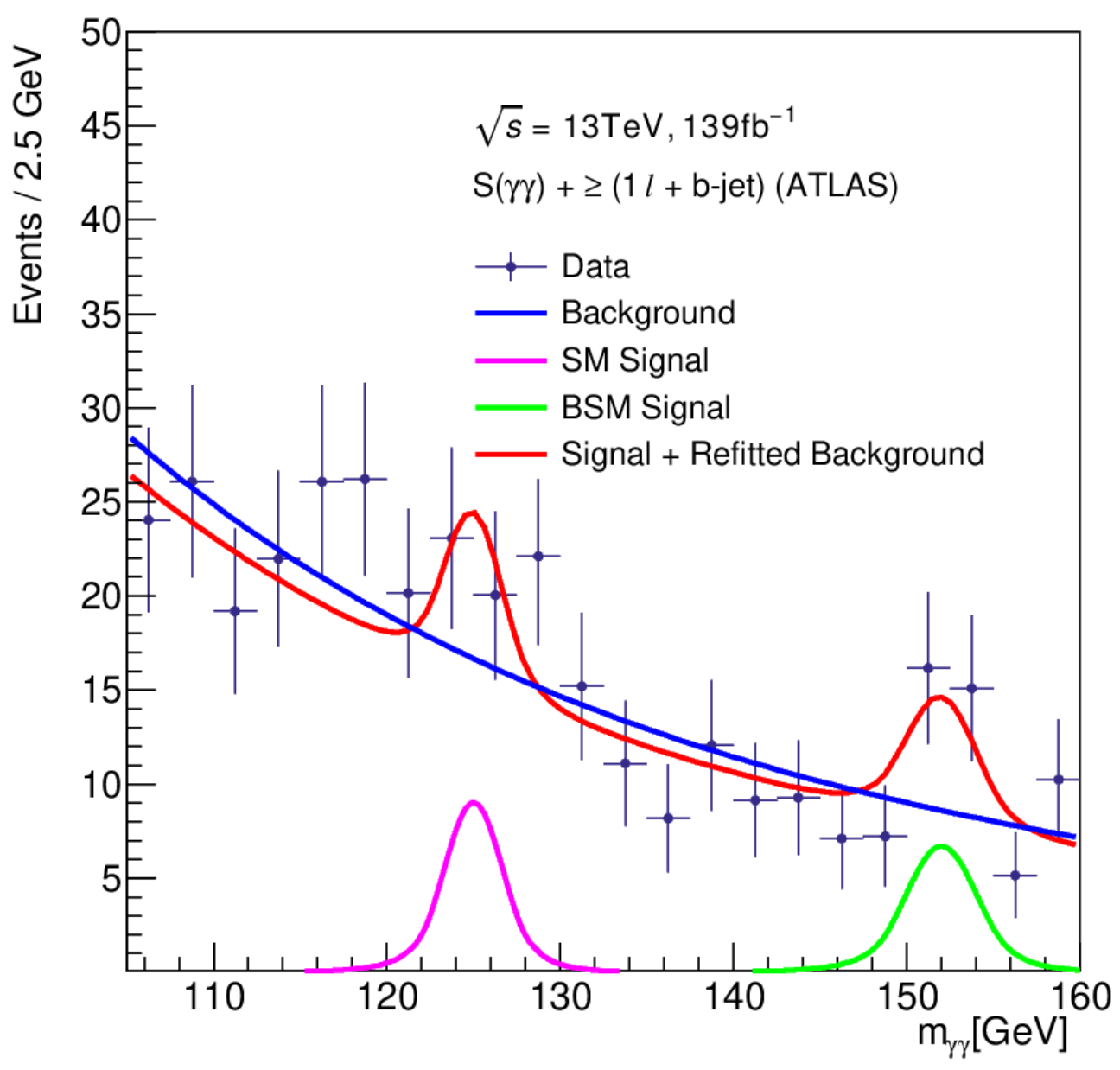}~~
    \includegraphics[width=0.4\textwidth]{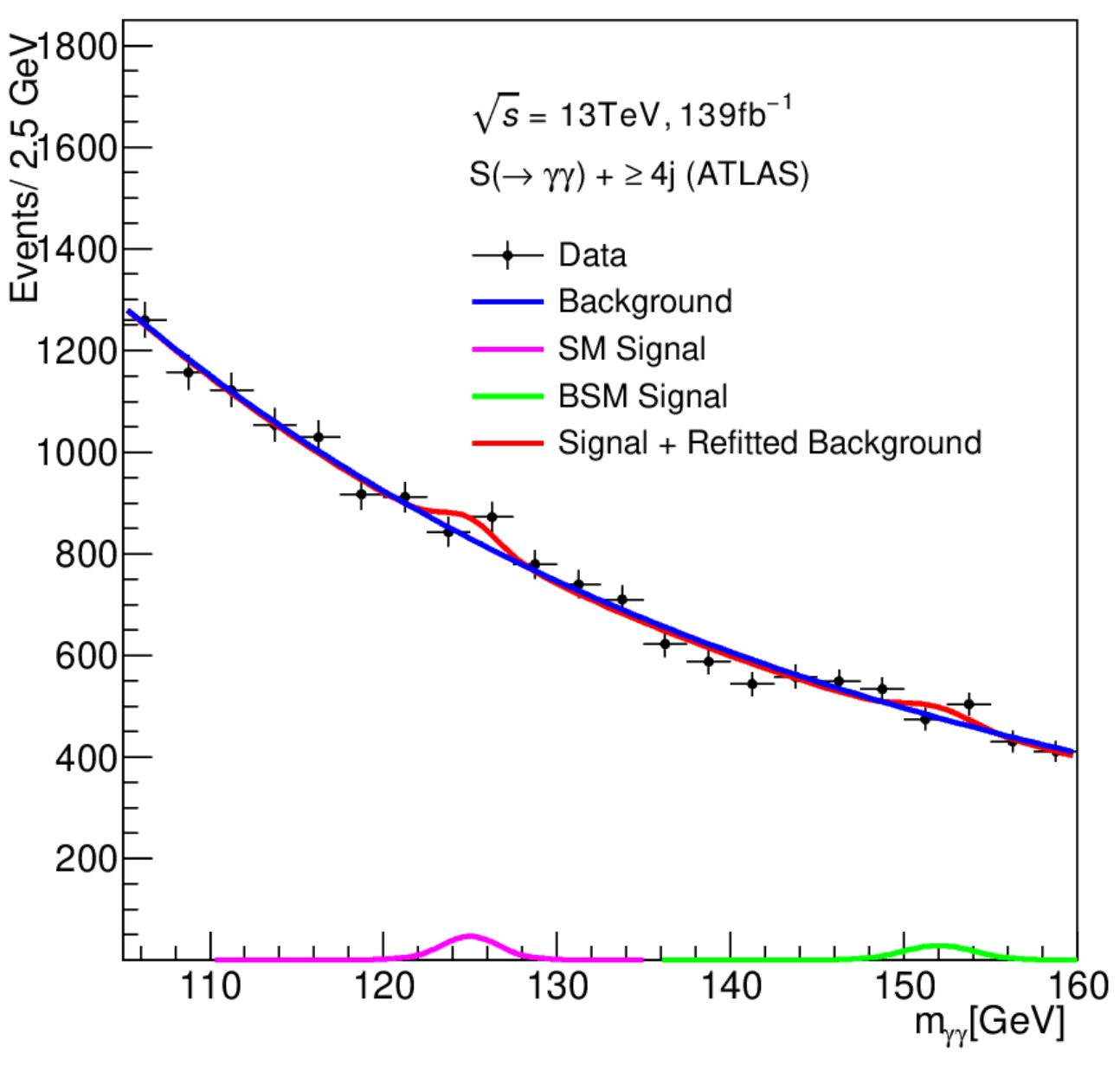}
    \caption{Diagrams showing the fit to background obtained within the SM, the SM Higgs signal, and the NP signal with the refitted background for $S (\to \gamma \gamma) + \geq (1\ell + 1 b$-jet) \cite{ATLAS:2023omk} and $S (\to \gamma \gamma) + \geq $ 4 jet \cite{ATLAS:2023omk}  categories. }
    \label{fig:invmass2}
\end{figure*}

\begin{figure*}[htbp!]
    \centering
    \includegraphics[width=0.4\textwidth]{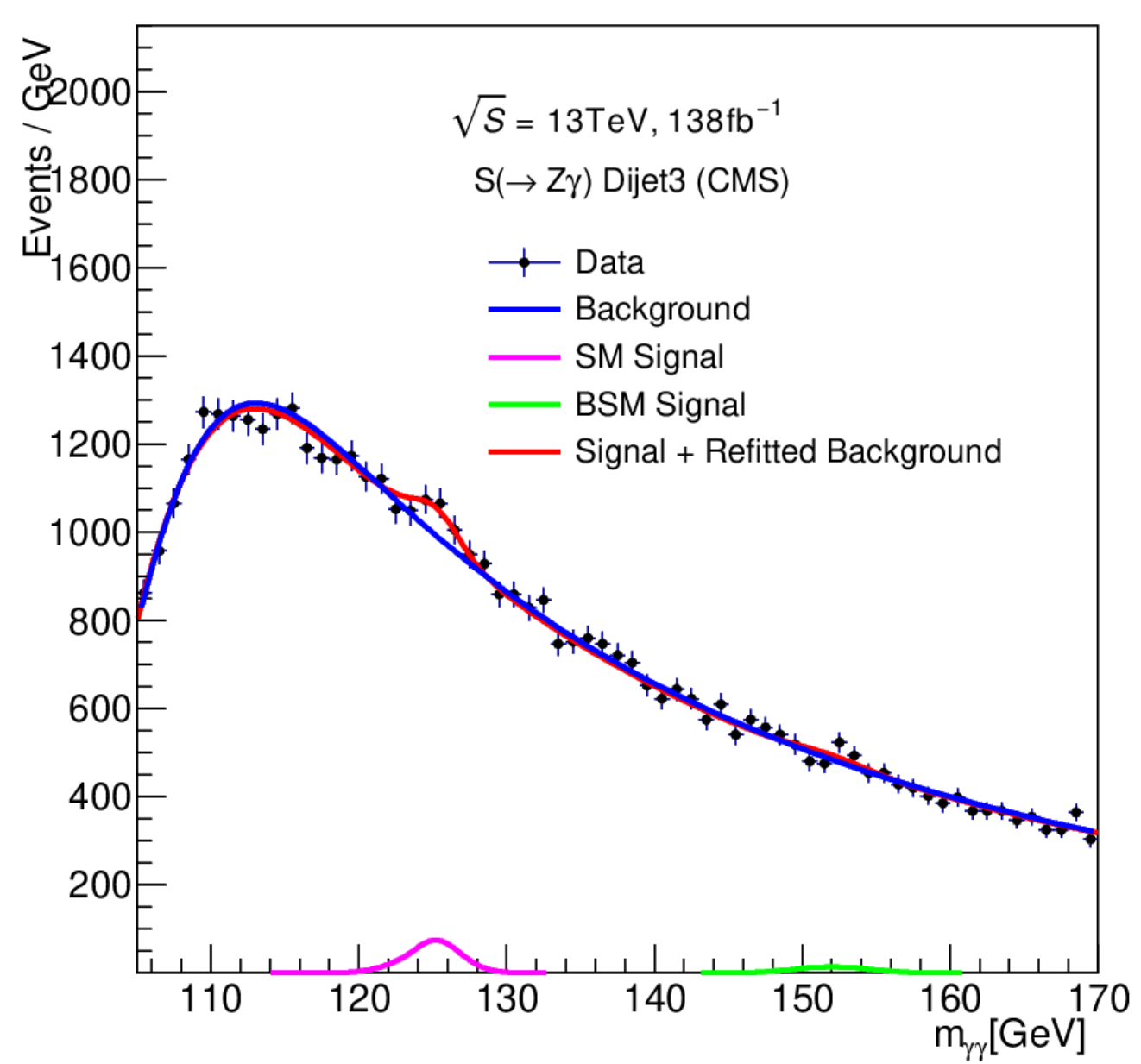}~~
    \includegraphics[width=0.4\textwidth]{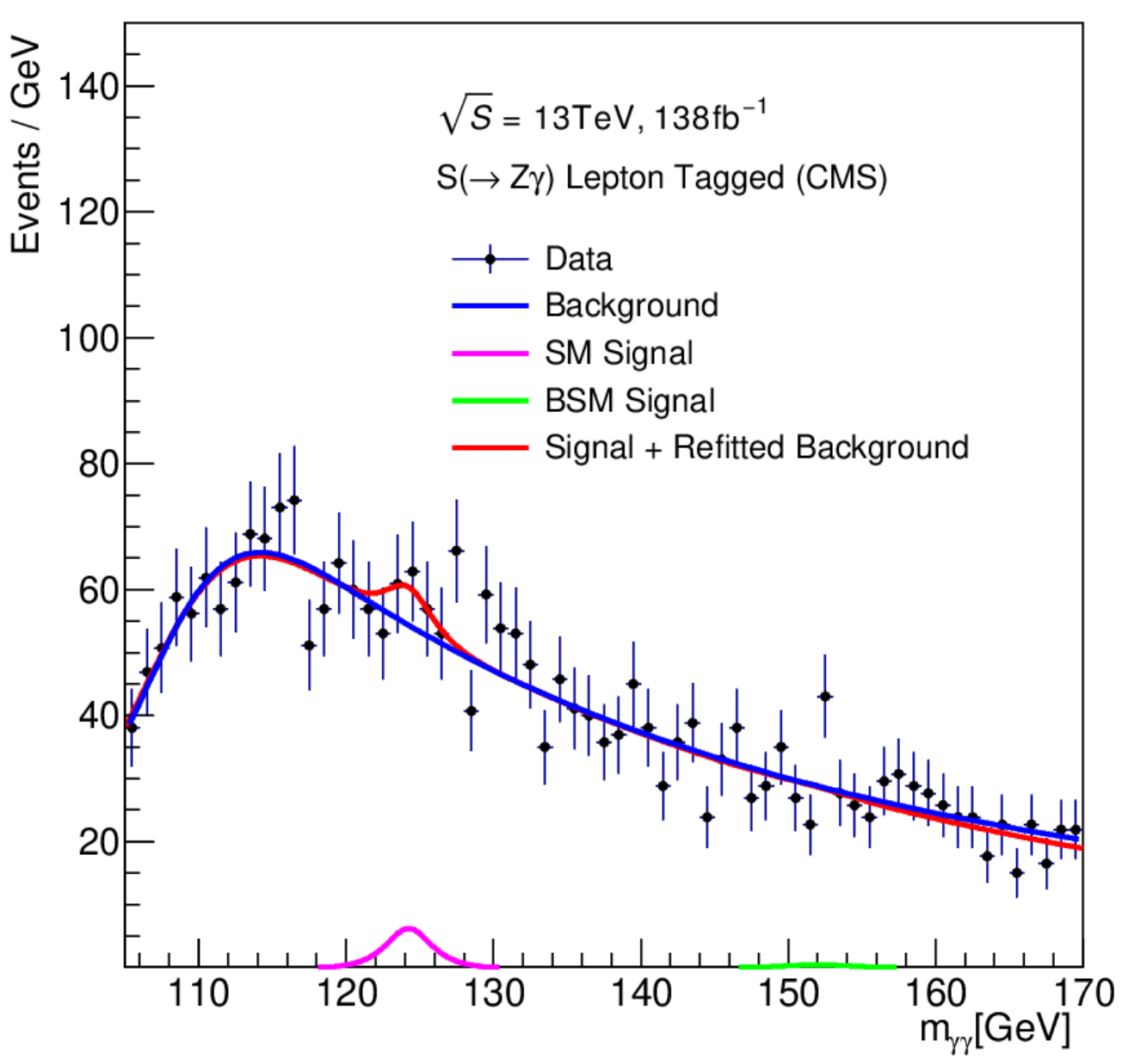}
    \caption{Diagrams showing the fit to background obtained within the SM, the SM Higgs signal, and the NP signal with the refitted background for two different $S(\rightarrow Z\gamma) $ categories from ref.~\cite{CMS:2022ahq} }
    \label{fig:invmass3}
\end{figure*}

\begin{figure*}[htbp!]
    \centering
    \includegraphics[width=0.7\linewidth]{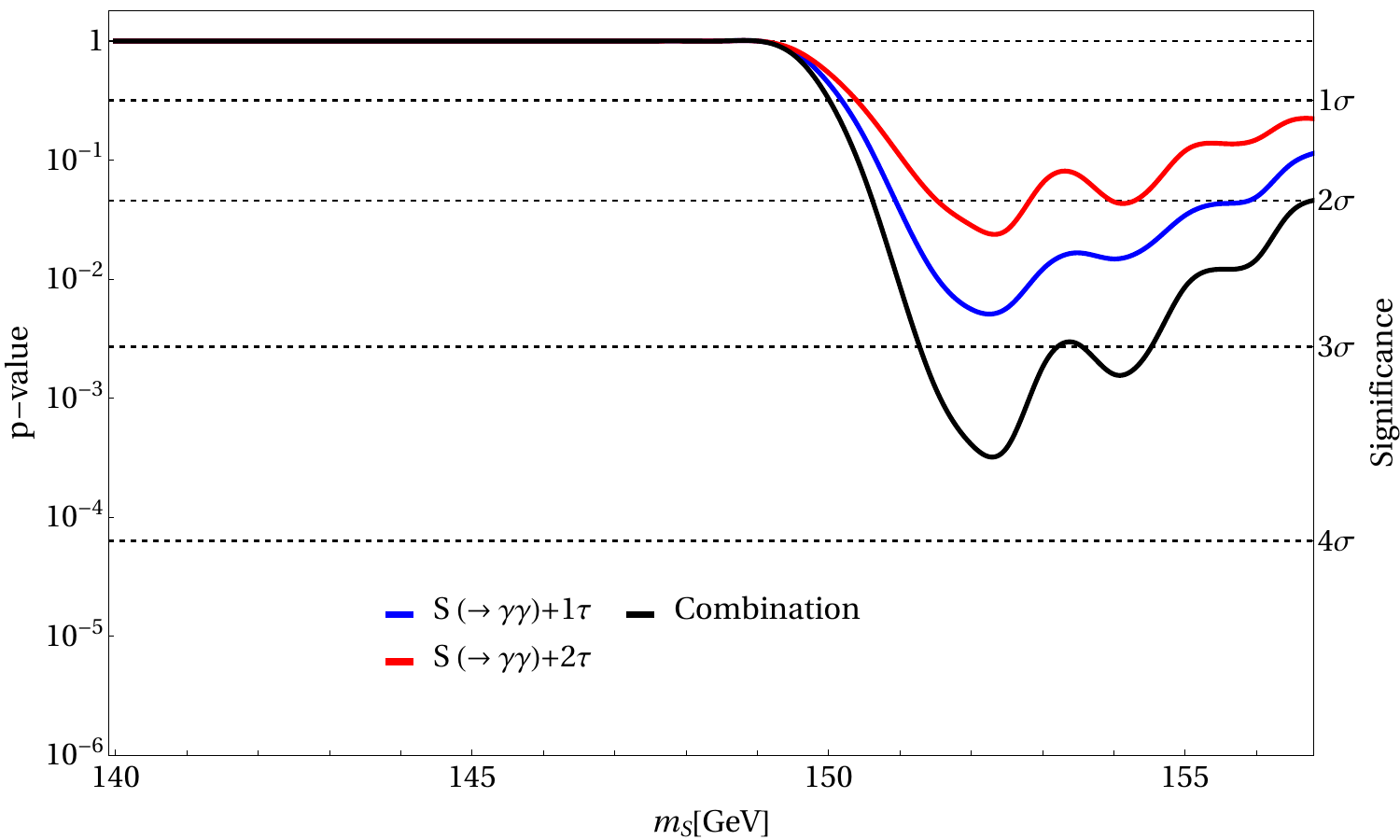}
    \caption{The $p$-values of the $S(\rightarrow \gamma \gamma) + 1 \tau$ and $S(\rightarrow \gamma \gamma) + 2 \tau$  channels with their combination and the corresponding local significance. The best fit is obtained around 152\,GeV with a local significance of 3.55$\sigma$. \label{fig:1tau2taucombined}}
\end{figure*}

%\section{Illustration of a Prominent Narrow Structure around 152 GeV}\label{secA3a}

\begin{figure*}[htbp!]
    \centering

    \includegraphics[width=0.9\linewidth]{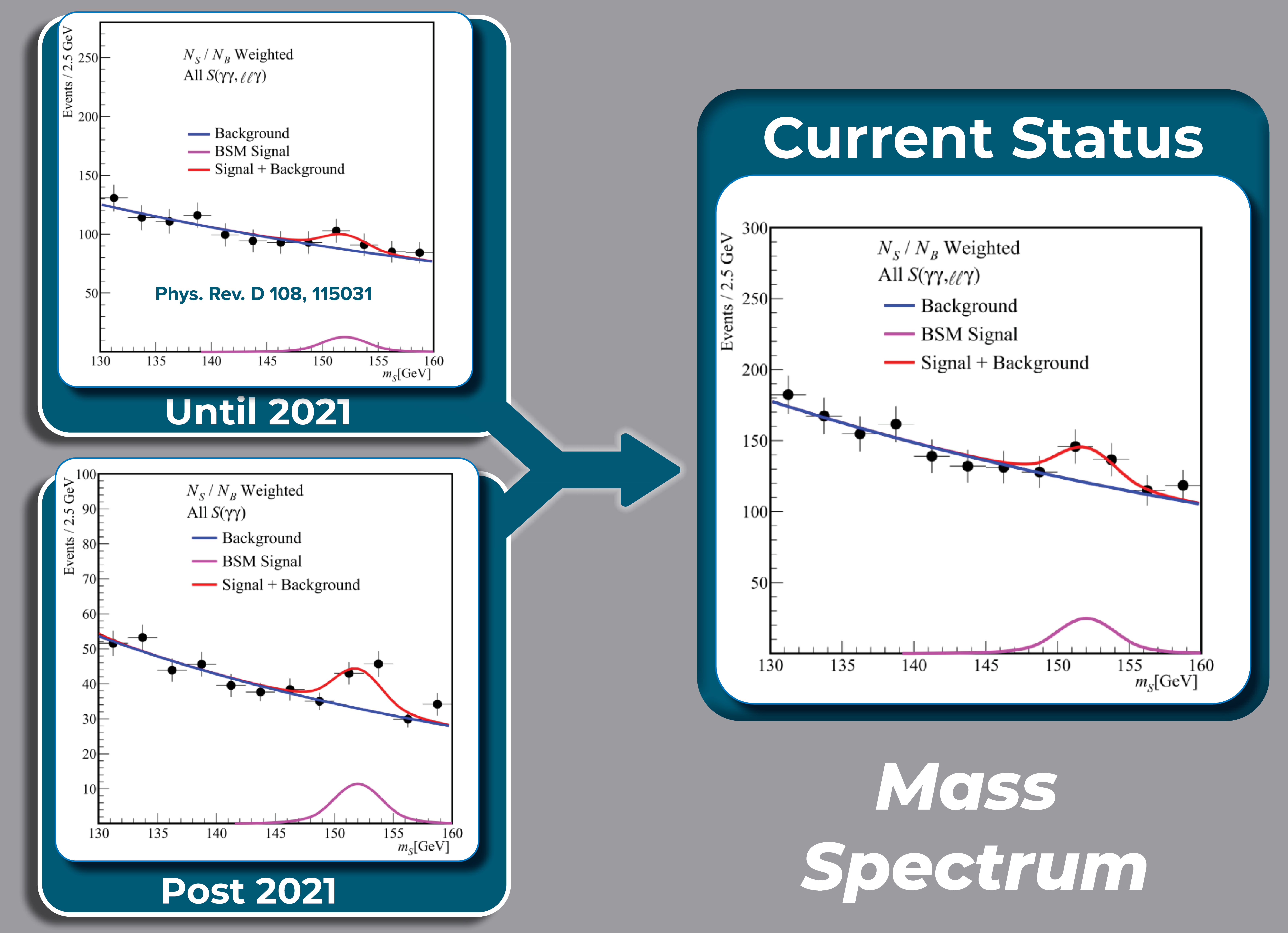}
    \caption{The illustrative representation of the $\gamma\gamma$ and $Z\gamma$ mass spectra (see text).}
    \label{massspectrum}
\end{figure*}

Now, we illustratively point towards a prominent narrow structure consistent with the detector resolution around 152 GeV.

Each graph in \autoref{massspectrum} combines the spectra from the LHC experiments for the same final states by re-weighting using signal-over-background yields ($N_S / N_B$).

Here, $N_S$ represents the signal yield within a mass window $\pm 2\sigma_{\text{res}}$ at the peak; $\sigma_{\text{res}}$ is the di-photon invariant mass resolution embedded into a Crystal Ball function used to model the resonance, and $N_B$ represents the corresponding background events within this range.

The mass spectra are shown for the data sets before (upper left) and after (lower left) 2021. The graph on the right shows the weighted addition of the data sets considered here. A prominent narrow structure consistent with the detector resolution around 152 GeV can be appreciated. However, it is important to note that this graph is not applicable for the significance calculation around 152 GeV.

\section{Fiducial Cross-sections of the 152 GeV Candidate}\label{secA4}

\begin{table}[htbp!]
    \centering
    \begin{tabular}{c|c}
        \toprule % \hline
       Channel  &  Efficiency ($\%$)  \\
       \midrule % \hline
        $\gamma \gamma$ & 60~\cite{ATLAS:2024lhu}\\
        %\hline
         $\tau$ & 80~\cite{ATLAS:2024lhu}\\
        %\hline
          $\ell (e,\mu)$ & 88~\cite{ATLAS:2019qmc, ATLAS:2020auj} \\
        %\hline
        $b$-jet & 40-70~\cite{Bhattacharya:2011pc}\\
                %\hline
        $Z(\to \ell \ell)\gamma$ & 20-31~\cite{CMS:2018myz, ATLAS:2020qcv}\\
        \bottomrule % \hline
    \end{tabular}
    \caption{The efficiency for different channels used in this analysis.}
    \label{tab:efficiency}
\end{table}

In this section, we tabulate the individual significance and the fiducial cross section $(\sigma_F)$ of each of the channels considered in this analysis. The cross-sections are measured in two ways: across the entire kinematic range and inside a fiducial phase space that is as near as possible to the experimental measurement range. The fiducial phase space is established using stable particles produced by Monte Carlo (MC) generators, which are then used to create reconstructed objects such as primary leptons, jets, and missing transverse momentum. The advantage of fiducial cross-section measurements is a significant reduction in the amount of the applied acceptance corrections, resulting in lower systematic errors.
The fiducial cross-section is defined as follows:
\begin{equation}
    \sigma_F = \frac{\text{No. of Signal events }(N_{\text{Sig}})}{\text{Efficiency}(\epsilon) \times \text{Integrated Luminosity}(\mathcal{L})}
    \label{fiducialcrosssection}
\end{equation}
Where $N_{sig}$ denotes the number of signal events, $\epsilon$ defines the efficiency of different decay modes (mentioned in~\autoref{tab:efficiency}), and $\mathcal{L}$ is the integrated Luminosity.

\begin{table}[htbp!]
    \begin{tabular}{|c|c|c|c|}
          \hline
       Channel  &  $N_{\text{Sig}}$ & Significance & $\sigma_F$ in fb\\
          \hline
        $S (\to \gamma \gamma) + \tau$  & 32.67 $\pm$ 11.49 & 3.56 & 0.49 $\pm$ 0.17\\
          \hline
        $S (\to \gamma \gamma) +  \ell$ &  10.72 $\pm$ 7.45 & 1.76 & 0.14 $\pm$ 0.08\\
          \hline
        $S (\to \gamma \gamma) +  2(\ell,\tau)$ &  0.43 $\pm$ 1.31 & 0.24 & 0.004 $\pm$ 0.01\\
          \hline
        $S (\to \gamma \gamma) + \geq 1\ell + 1 b$-jet  & 13.30 $\pm$ 6.16 & 2.41 & 0.24 $\pm$ 0.11\\
         \hline
        $S (\to \gamma \gamma) + \geq $ 4 jet  & 57.98 $\pm$ 35.30 & 1.46 & 0.80 $\pm$ 0.42\\
         \hline
        $S(\to Z(\to \ell^+ \ell^-)\gamma) + jj$  & 88.61 $\pm$ 58.75 & 1.24 & 2.16 $\pm$ 1.37\\
         \hline
        $S(\to Z(\to \ell^+ \ell^-)\gamma) + 1\ell$  & 2.49 $\pm$ 15.48 & 0.14 & 0.12 $\pm$ 0.41 \\
         \hline
        $S (\to \gamma \gamma) $ + $b-$jet  & 14.07 $\pm$ 6.64 & 2.49 &  0.22 $\pm$ 0.10\\
         \hline
        $S(\to \gamma\gamma) + E^T_{\rm miss}$ (Low)   & 21.10 $\pm$ 7.54 & 2.82 & 0.25 $\pm$ 0.09 \\
         \hline
        $S(\to \gamma\gamma) + E^T_{\rm miss}$ (High)  & 8.19 $\pm$ 3.68 & 2.53 & 0.10 $\pm$ 0.04\\
         \hline
         $S(\to \gamma\gamma) + W,Z$ (CMS)   & 15.68 $\pm$ 14.75 & 0.96 & 0.19 $\pm$ 0.18\\
         \hline
        $S(\to \gamma\gamma) + W,Z$ (ATLAS)   & 4.12 $\pm$ 12.55 & 0.35 & 0.06 $\pm$ 0.17\\
         \hline
    \end{tabular}
    \caption{Fiducial Cross section for different channels used in search of the narrow resonance.}
    \label{tab:Fiducial_crosssection}
\end{table}

\newpage
\end{appendices}


\bibliography{sn-bibliography}% common bib file

\end{document}